# Aligning free surface properties in time-varying hydraulic jumps

Rui LI[1], Kristen D. SPLINTER[2] and Stefan FELDER[3]

[1] PhD Candidate, UNSW Sydney, Water Research Laboratory, School of Civil and Environmental Engineering, 110 King St, Manly Vale, NSW, 2093, Australia
ORCID: 0000-0002-7387-8004

[2] Senior Lecturer, UNSW Sydney, Water Research Laboratory, School of Civil and Environmental Engineering, 110 King St, Manly Vale, NSW, 2093, Australia
ORCID: 0000-0002-0082-8444

[3] Senior Lecturer, UNSW Sydney, Water Research Laboratory, School of Civil and Environmental Engineering, 110 King St, Manly Vale, NSW, 2093, Australia
+61 (2) 8071 9861; s.felder@unsw.edu.au (corresponding author)
ORCID: 0000-0003-1079-6658

**Abstract**

Hydraulic jumps occur commonly in natural channels and energy dissipation systems of hydraulic structures in the violent transition from supercritical to subcritical flows. They are characterised by large flow aeration, high turbulence and strong fluctuations of the free surface and the jump toe. For free surface measurements, fast-sampling, fixed-point instruments such as acoustic displacement meters (ADMs) and wire gauges (WGs) are commonly used, while LIDAR technology is a relatively new method for recording instantaneous free surface motions of aerated flows. While each of these instruments has been shown previously to provide reasonable results for basic and advanced free surface properties, differences between instruments and experiments remain unexplained. To systematically analyse these differences, simultaneous laboratory experiments of aerated hydraulic jumps were conducted. Good agreement between the three instruments was obtained for basic free surface properties including elevations, fluctuations, skewness, kurtosis, and frequencies, as well as advanced free surface properties such as integral time and length scales. These new results indicate that any of these instruments can be used for the recording of free surface properties albeit the integration limit for free surface scales must be considered. A key finding of this research was that differences between repeated experiments as well as previous studies were observed when using the visual jump toe for alignment. However, this bias could be resolved by using the mean jump toe location recorded with the LIDAR. Therefore, future studies should simultaneously measure the instantaneous jump toe to provide more consistent results across studies.

# 1. Introduction

Hydraulic jumps are commonly observed in natural open channels or as energy dissipators in stilling basins [1,2]. They occur at the violent transition from supercritical to subcritical flows and are associated with large air entrainment, turbulence and energy dissipation [3–5]. As such, hydraulic jumps have been extensively researched, including their free surface properties (Table 1). As shown in Figure 1, air is entrained locally at the impingement point (jump toe) and continuously entrained and detrained along the roughened surface of the jump roller [6–8]. The air-water interface is characterized by entrained bubbles, water droplets and air pockets trapped in the free surface roughness [7]. The flow is highly three-dimensional, with both fast and slow jump toe motions linked to internal vortex pairing and shedding [8–11]. The complex three-dimensional flow structure combined with the intense longitudinal motions raise significant challenges for measuring the flow properties, particularly from point source measurements as shown in Figure 1.

Previous experiments have mainly focused on the internal flow structures including air-water flow properties using point measurements with phase-detection intrusive probes (e.g. [12–14]), internal pressures using a point source pressure sensor [8,15] and internal turbulence characteristics using both point source measurements (e.g. [10,16,17]) and video-based detection (e.g. [18]).

The free surface features of hydraulic jumps have also been extensively studied since they provide important insight into fundamental understanding of the flow processes and the design of hydraulic structures. Previous research has predominately used point-source measurements (Table 1). For example, pointer gauges (e.g. [19,20]) have been applied to estimate time-averaged free surface profiles, while profiling with point-source phase-detection intrusive probes have been used to provide an indirect air-water flow measurement of the time-averaged free surface elevation (e.g. [8,21]). Advancements in instrumentation, including the use of wire gauges (WG), acoustic displacement meters (ADMs), LIDAR, and high-speed cameras have enabled instantaneous measurements of free surface properties including fluctuations, characteristic frequencies and free surface integral scales. Image-based techniques can also be applied for free surface measurements to record continuous free surface profiles at the sidewall [9,10,18]. However, the sidewall dampens the three-dimensional hydraulic jump motions, and the free surface differs from the centreline data especially in wide channels [22].

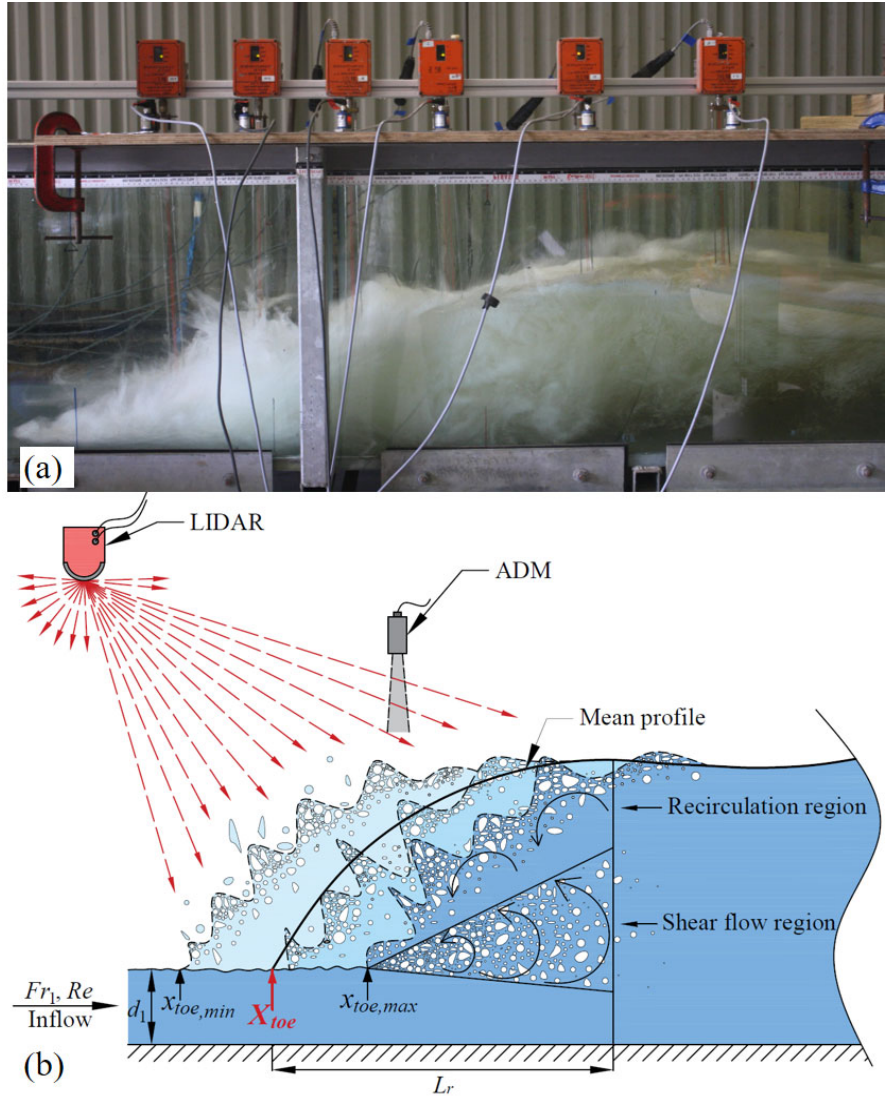

Figure 1. Side view of an aerated hydraulic jump: (a) Photo of present study with $d_1$ = 0.028 m, $Fr_1$ = 8, $Re$ = 1.2×10$^5$; ADMs (foreground) and WGs (orange boxes with stems in background) installed at a fixed location above the flume; (b) conceptual sketch of hydraulic jump characteristics including longitudinal movements around the mean jump toe $X_{toe}$ with minimum and maximum extent of jump toe positions $x_{toe,min}$ and $x_{toe,max}$ respectively.

Table 1. Relevant experimental studies of free surface properties in hydraulic jumps using ADMs, WGs and LIDAR ($d_1$ = inflow depth, $Fr_1$ = inflow Froude number, $Re$ = Reynolds number, $W$ = channel width).

| Reference | $d_1$ [m] | $Fr_1$ [-] | $Re$ [-] | $W$ [m] | Inflow Condition | Instrument (Sampling Frequency) | No. of simultaneous measurement locations | Sampling duration [s] | Elevation | Fluctuations | Frequency | Time and length scales |
|---|---|---|---|---|---|---|---|---|---|---|---|---|
| Mouaze et al. [23], Murzyn et al. [24] | 0.021-0.059 | 1.9-4.8 | $3.3\times10^4 - 8.9\times10^4$ | 0.3 | Partially developed | WG (128 Hz) | N/A | 5 | x | x | | x |
| Murzyn and Chanson [25] | 0.018 | 3.1-8.5 | $2.4\times10^4 - 6.4\times10^4$ | 0.5 | Partially developed | ADM (50 Hz) | 6 | 600 | x | x | x | |
| Chachereau and Chanson [26] | 0.038 - 0.045 | 2.4 - 5.1 | $6.6\times10^4 - 1.3\times10^5$ | 0.5 | Partially developed | ADM (50 Hz) | 7 | 600 / 60 | x | x | x | x |
| Nóbrega et al. [27] | 0.027 | 2.4 | $3.3\times10^4$ | 0.48 | N/A | ADM (25 Hz) | 1 | 120 | x | | | |
| Wang and Chanson [28], Wang et al. [8] | 0.012-0.054 | 3.8-10 | $2.1\times10^4 - 1.6\times10^5$ | 0.5 | Partially developed | ADM (50 Hz) | 5 | 540 | x | x | x | |
| Montano et al. [11] | 0.032 – 0.154 | 2.1 – 4.7 | $8.4\times10^4 - 3.9\times10^5$ | 0.5 / 0.6 | Fully developed | LIDAR (35 Hz) | 120-195 | 1800 | x | x | x | |
| Montano and Felder [29] | 0.02 – 0.046 | 3.6 – 10 | $6.2\times10^4 - 1.2\times10^5$ | 0.6 | Fully developed | LIDAR (35 Hz) | - | 1800 | x | x | | x |
| Stojnic et al. [30] | 0.048-0.066 | 6.2-13.1 | $2.0\times10^5 - 3.6\times10^5$ | 0.5 | Fully developed and aerated | ADM (12.5 Hz) | 1 | 328 | x | x | | |
| Present Study | 0.041 0.034 0.028 | 3.5 5 8 | $9.2\times10^4$ $1.0\times10^5$ $1.2\times10^5$ | 0.6 | Fully developed | LIDAR (35 Hz) | 135-180 | 600 -1800 | x | x | x | x |
| | | | | | | ADM (100 Hz) | 6 | | | | | |
| | | | | | | WG (100 Hz) | 6 | | | | | |

Table 1 summarises experimental studies of free surface features in hydraulic jumps focussing on fast-sampling instrumentation applied along the channel centreline. While WGs and ADMs can record instantaneous free surface motions at a single fixed point per instrument, LIDAR technology allows the simultaneous and continuous recording of free surface motions with high spatial resolution along the entire hydraulic jump.

Montano *et al.* [11] showed that LIDAR measurements of basic free surface properties including time-averaged free surface profiles, fluctuations and characteristic frequencies are in general agreement with previous free surface data recorded with ADMs, WGs and pointer gauges. However, distinct differences were observed in basic free surface properties between previous studies in terms of free surface elevations and fluctuations [11]. In addition, Montano and Felder [29] also found discrepancies in reported free surface integral time and length scales between comparable studies using different instrumentation and post-processing methods.

At present, it is unclear what might cause the reported differences in free surface properties between comparable hydraulic jumps. For example, the inflow conditions and boundary layer development upstream of hydraulic jumps can affect the internal motions and flow aeration [13,31] as well as the free surface properties [32]. In addition, the jump toe within a hydraulic jump varies considerably in both space and time [9,11,28,29] making it quite challenging to robustly define a mean jump toe position $X_{toe}$. Traditionally $X_{toe}$ has been determined visually and the comparison of free surface properties between different studies that used the visually observed mean jump toe $X_{toe,visual}$ as the reference frame may be one source of variability between previous measurements. Different experimental facilities, types of instruments, sampling time and post-processing methods may also contribute to observed differences.

Therefore, the objectives of the present study are to (1) provide an explanation of previously reported differences in basic and advanced free surface properties in fully aerated hydraulic jumps, (2) to identify the contributions of instrumentation, data processing and frame of reference on this and to (3) provide a robust method to collapse data into a single frame of reference for improved comparison between studies. To achieve these objectives, simultaneous measurements using LIDAR, ADMs and WGs were conducted for the first time in a controlled laboratory environment. Fully aerated hydraulic jumps were measured with identical experimental environments, including inflow conditions, facilities, measurement locations, reference frames and sampling parameters to achieve the objectives. Comparative analyses of basic free surface properties including mean surface elevations, fluctuations and frequencies, as well as advanced parameters including integral free surface time and length scales were performed to understand the differences in free surface properties measured by different instruments showing that a close alignment of all basic properties can be achieved through the use of the mean jump toe determined by the LIDAR $X_{toe,LIDAR}$ as the reference frame.

## 2. Methodology
### 2.1 Experimental setup and flow conditions

New experiments were conducted in a flume of 40 m length and 0.6 m width at the UNSW Water Research Laboratory. Supercritical flows entered the flume underneath a sluice gate with an upstream rounded corner (Figure 2). The flow was controlled with an ABB WaterMaster® FET100 electromagnetic flowmeter with an accuracy of ±0.4% of the flow rate. More details on the experimental setup can be found in Montano [21]. The experiments comprised three different hydraulic jumps with fully developed inflow conditions with Froude numbers $Fr_1 = \frac{v_1}{\sqrt{gd_1}}$ = 3.5, 5 and 8, corresponding to Reynolds numbers $Re = q/v$ = 9.2×10$^4$, 1×10$^5$ and 1.2×10$^5$, where $v$ is the kinematic viscosity of water, and $q$ the discharge per unit width. The inflow depth $d_1$ was measured 6 times with a pointer gauge at the location of the jump toe and the average value was taken. To ensure small variations in inflow conditions or jump toe positions did not influence the results, the free surface was simultaneously sampled with LIDAR, ADMs and WGs (Figure 2).

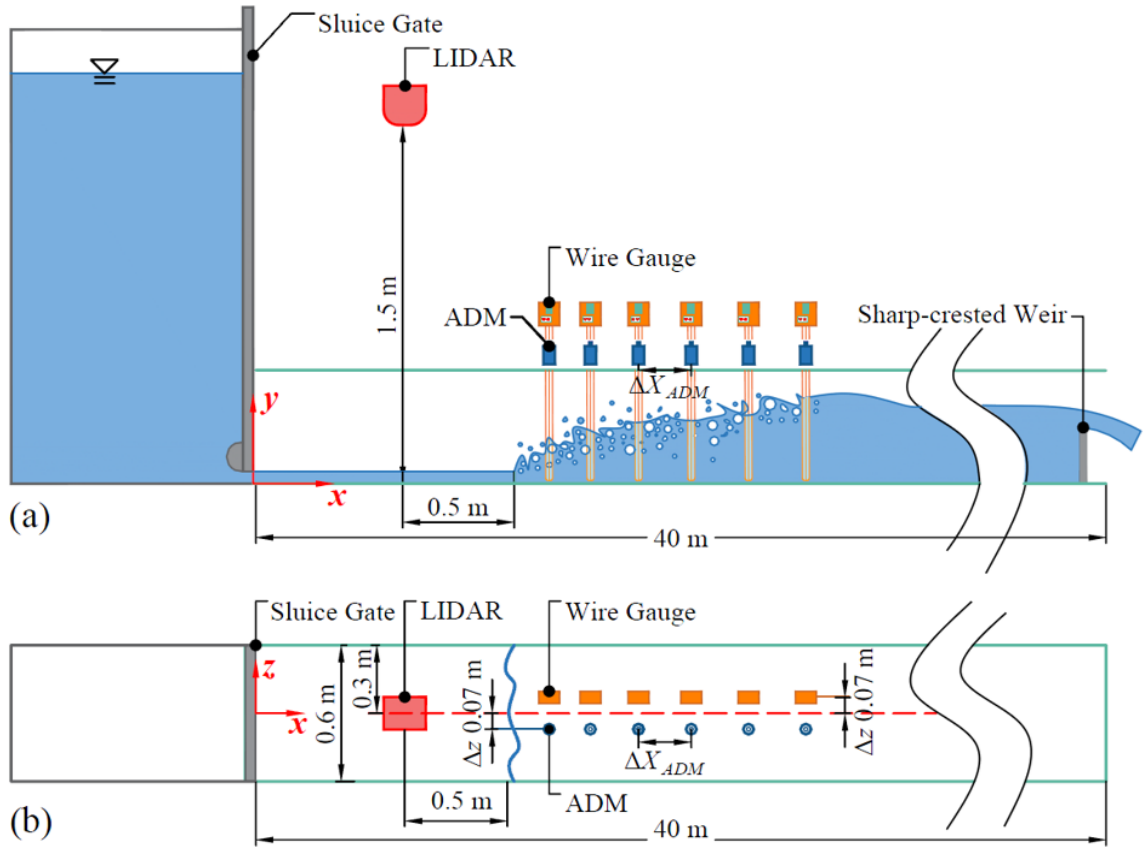

Figure 2. Experimental setup and positioning of instrumentation (not to scale): (a) Side view; (b) Top view.

The LIDAR was a SICK LMS511 sampling with a frequency of 35 Hz and an angular resolution of 0.25°, which was consistent with Montano et al. [11] and Montano and Felder [29]. The LIDAR transmitted laser pulses and distances were calculated based on the Time-Of-Flight principle [33,34]. The LIDAR raw data were recorded using the SOPAS ET software of SICK [34]. The LIDAR was positioned 1.5 m above the channel bed and approximately 0.5 m upstream of the jump toe to provide the best possible perspective for the free surface measurements [35]. As per the manufacturer manual [34], the LIDAR had systematic and statistical errors of ±25 mm and ±6 mm, respectively, and a laser beam width of < 20 mm albeit laboratory tests with spacers suggested a beam width as low as 5 mm for the present test conditions.

Six ADMs (Microsonic[TM] Mic+35/IU/TC) with an accuracy of ±1% (temperature drift compensated) and a vertical resolution of 0.17 mm for the selected operating range of 65 to 600 mm were used. A cylindrical hollow extension of 4 cm was attached to the ADMs to protect the sensor head from water splashes as well as interference of adjacent sensors as suggested by Kramer and Chanson [36]. As per the manufacturer, the spot size of the ADM within the present experimental setup was ca. 100 mm. However, considering the cylindrical extension that limited the signal spreading in the upper part and according to Zhang *et al*. [37], the actual spot size was more likely in the range of 50 to 80 mm. The ADMs recorded the first return signal. The relatively large spot size of ADMs may result in erroneous data capture due to water splashing, signal interference by adjacent sensors [26,28,38], or slope effects if the free surface slope is greater than 13.5° [37]. In aerated hydraulic jumps strong free surface non-stationarities, splashing and slope affect the data quality recorded by ADMs.

In addition, six capacitance WGs (Manly Hydraulics Laboratory, Sydney) were used. The WGs consisted of a dielectric coated wire (Ø = 0.2 mm) of 200 mm length supported by a metal frame (Figure 1). Depending upon the length of wire immersed in the water, the resistivity of the WG provides a measure of the flow depth. While little information is known about accuracies of WGs in hydraulic jumps, Mouaze *et al*. [23] and Murzyn *et al*. [24] suggested that strong turbulence and high aeration may introduce uncertainties. All WGs were calibrated in clear still water and any effects of the strong aeration on the WG measurements could not be assessed.

The raw voltage signals of ADMs and WGs were recorded digitally with LabVIEW on the same computer used for LIDAR measurements and the internal computer time was used for synchronisation. ADMs and WGs were simultaneously sampled at 100 Hz to minimize aliasing distortion [37]. All instruments were warmed up for at least 1 hour before experiments to eliminate signal drifting. In all present experiments, no filtering was applied for ADMs and WGs during data acquisition and real time data were visually monitored. If the raw signal of any of the ADMs indicated a flat signal (corresponding to direct water impact), the recordings of all instruments were terminated, and the measurements repeated. The recorded raw data of LIDAR and ADMs/WGs did not have the identical start and end time since they were sampled with different acquisition software and were therefore manually trimmed to the same start and end time using the computer clock.

Figure 2 shows the experimental setup with all instruments. Measurements were conducted along the centreline as well as along two transects with a transverse offset of $\Delta z = \pm 0.07$ m (Figure 2b). To ensure that this small offset did not adversely influence the recorded signal with respect to comparing simultaneous instruments, a series of preliminary tests using the LIDAR and ADM were conducted along the three cross-sections (see Supplementary Material S1). These tests confirmed minimal transverse variations in free surface properties within the central part of the hydraulic jump.

The main experiments were conducted in three stages. During the first two stages basic free surface features were simultaneously recorded along the hydraulic jump with the different instruments. Data were recorded for six to seven repeated runs for each Froude number with sampling durations between 10 and 30 minutes. In the first stage, free surface measurements were simultaneously conducted with ADMs and LIDAR in the centreline or offset to either side by $\Delta z = \pm 0.07$ m. Experiments for each flow condition were repeated shifting LIDAR and ADMs between the three cross-sections ($\Delta z = 0, \pm 0.07$ m). The longitudinal distance between two consecutive ADM sensors, $\Delta X_{ADM}$, was $0.086 \leq \Delta X_{ADM} \leq 0.131$ m, $0.120 \leq \Delta X_{ADM} \leq 0.200$ m and $0.196 \leq \Delta X_{ADM} \leq 0.308$ m for $Fr_1 = 3.5$, 5.0 and 8.0, respectively. In the second stage of experiments, free surface measurements were recorded simultaneously with the LIDAR in the centreline and the ADMs and WGs at either side of the centreline (Figure 1 and 2). The longitudinal measurement locations of ADMs and WGs were identical as in the first stage.

In the third stage of experiments the free surface integral time and length scales were measured. This was done by simultaneously measuring the free surface with LIDAR, ADMs and WGs at distinct spacing intervals. For this set of experiments, the spacing of ADMs was $\Delta X_{ADM} = 47.5$ mm for the first five sensors and 95 mm for the last sensor. The distance between consecutive WGs was consistently 95 mm due to the sensors' electronic box limiting shorter spacing (Figure 1). For each flow condition, the array of ADMs and WGs was placed at one of the six ADM positions from the first and second stages. At each location, all sensors were sampled simultaneously for at least 15 minutes.

*2.2 Post-processing of raw data*

All signals were post-processed in MATLAB following the methodology shown in Figure 3. For the LIDAR, the raw data (distance and angle) were first translated into a cartesian coordinate system of elevations $y$ and distances $x$ along the flume using the recorded channel bed without water as the reference elevation. Based upon a detailed sensitivity analysis of filtering (See Supplementary Material S2), the LIDAR data were not filtered along the length of the jump roller $L_r$ [35]. Downstream of $L_r$, LIDAR data were filtered using 3 standard deviations of 12 neighbourhood points in the space domain and 4 standard deviations of 12 neighbourhood points in the time domain to remove outliers where insufficient aeration prevented the LIDAR from consistently returning the free surface [35]. The data quality was overall high, with less than 3% of non-detected or filtered LIDAR data in the conjugate depth region downstream of the jump roller for all experiments. In the next post-processing step, the instantaneous jump toe position $x_{toe}$ was automatically determined based on LIDAR measurements at each time step. Two data points closest to the inflow depth $d_1$ were determined by scanning from both the upstream and downstream directions. $x_{toe}$ was determined based on interpolation of the two data points to obtain the estimate of inflow depth $d_1$. Note that additional analysis with manual jump toe detection and machine learning algorithms suggested that this method was consistent for all flow conditions. The mean jump toe position derived from the LIDAR data $X_{toe.LIDAR}$ relative to the start of the sluice gate were calculated for each experiment. A mean jump toe was also visually estimated during each experiment $X_{toe.visual}$. In the fourth post-processing step, signals upstream of the jump toe with depth smaller than inflow depth $y < d_1$ due to signal penetration in the supercritical flow region were removed. These data were replaced with NaN at each time step to not bias the statistics during signal processing close to the jump toe. Note that previous LIDAR studies (e.g. [11,29], see Table 1) have replaced depths below the inflow depth with $d_1$. The number of unique data points recorded along the jump roller at every scan was approximately 75, 95 and 122 for hydraulic jumps with $Fr_1$ = 3.5, 5 and 8, respectively. In the final step of post-processing, the LIDAR data were then interpolated with a constant longitudinal distance with spacing ranging between 8 mm ($Fr_1$ = 3.5) and 12 mm ($Fr_1$ = 8). This final step ensured that all statistical calculations (mean, standard deviation, etc) were done on the same grid for each flow condition.

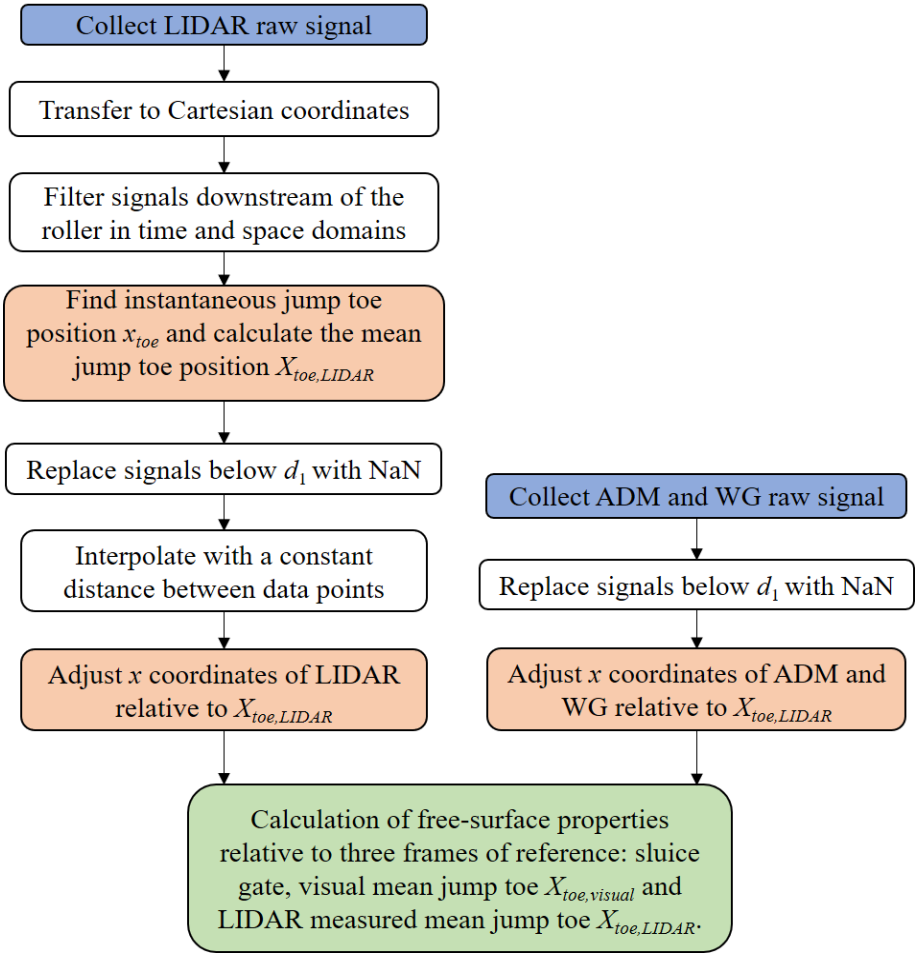

Figure 3. Post-processing steps of LIDAR, ADM, and WG data.

The post-processing framework of the raw ADM and WG data is also shown in Figure 3. A manual check of raw data (Figure 4) showed very few outliers in contrast to previous reports by Wang [39]. Using the work of Valero *et al.* [40] as a guide, different post-processing filtering methods were tested for ADM and WG data comprising simple cut-off thresholds based on standard deviations and percentiles, as well as the elliptical bound filter based on sampled flow elevations and its derivative (vertical velocity) using the method of Goring and Nikora [41]. A detailed discussion of the tested filtering methods (including down-sampling) and the effects on the free surface properties is presented in Supplementary Material S2. For the results presented here, no filtering was applied to avoid potential removal of meaningful data. During the post-processing, the ADM and WG data below the inflow depth were removed and replaced with NaN. This only had impact on the first ADM/WG sensor that was affected by jump toe oscillations and could be as high as 40% for $Fr_1 = 8$ due to the positioning of the first sensors close to $X_{toe.visual}$.

For all instruments, three different frames of reference were considered. The first was the absolute frame of reference, relative to the sluice gate corresponding to $x = 0$. The second reference frame was relative to the mean jump toe estimated visually $X_{toe.visual}$. This was used prior to the recordings to position the hydraulic jump at the same location within the flume for repeated experiments. The third frame of reference was relative to the mean jump toe recorded by the LIDAR $X_{toe,LIDAR}$.

Figure 4 shows a 20 second segment of raw unfiltered data for the three measurement devices as well as the probability mass functions (PMFs) for the entire sampling duration (1380 s) of a single test. All raw time series data showed strong fluctuations in free surface elevations of similar magnitude and with similar patterns. The raw data for the ADM (Figure 4b) and the WG (Figure 4c) had a more continuous signal compared to the LIDAR (Figure 4Figure 4a). This appears to be linked with the higher sampling frequency for the ADMs and WGs (100 Hz) compared to the LIDAR (35 Hz) as well as spot size for the ADMs and some smoothing due to wetting and drying times of the intrusive WGs. The PMFs of all three instruments over all flow conditions showed similar distributions independent of the measurement locations and flow conditions. As detailed above, while the signals were recorded simultaneously, they were not measured at the identical transverse location ($\Delta z = 0, \pm 0.07$ m). As a result, the aim of the present study was not to compare the instantaneous free surface signals but to focus upon statistical properties of the free surface. The simultaneous recordings were however essential to have the same frame of reference and to remove any small differences in inflow conditions and human bias in determination of $X_{toe,visual}$. The agreement between the instantaneous signals was indirectly assessed in a cross-correlation analysis of the simultaneously sampled time series data for the three instruments (see Supplementary Materials S1).

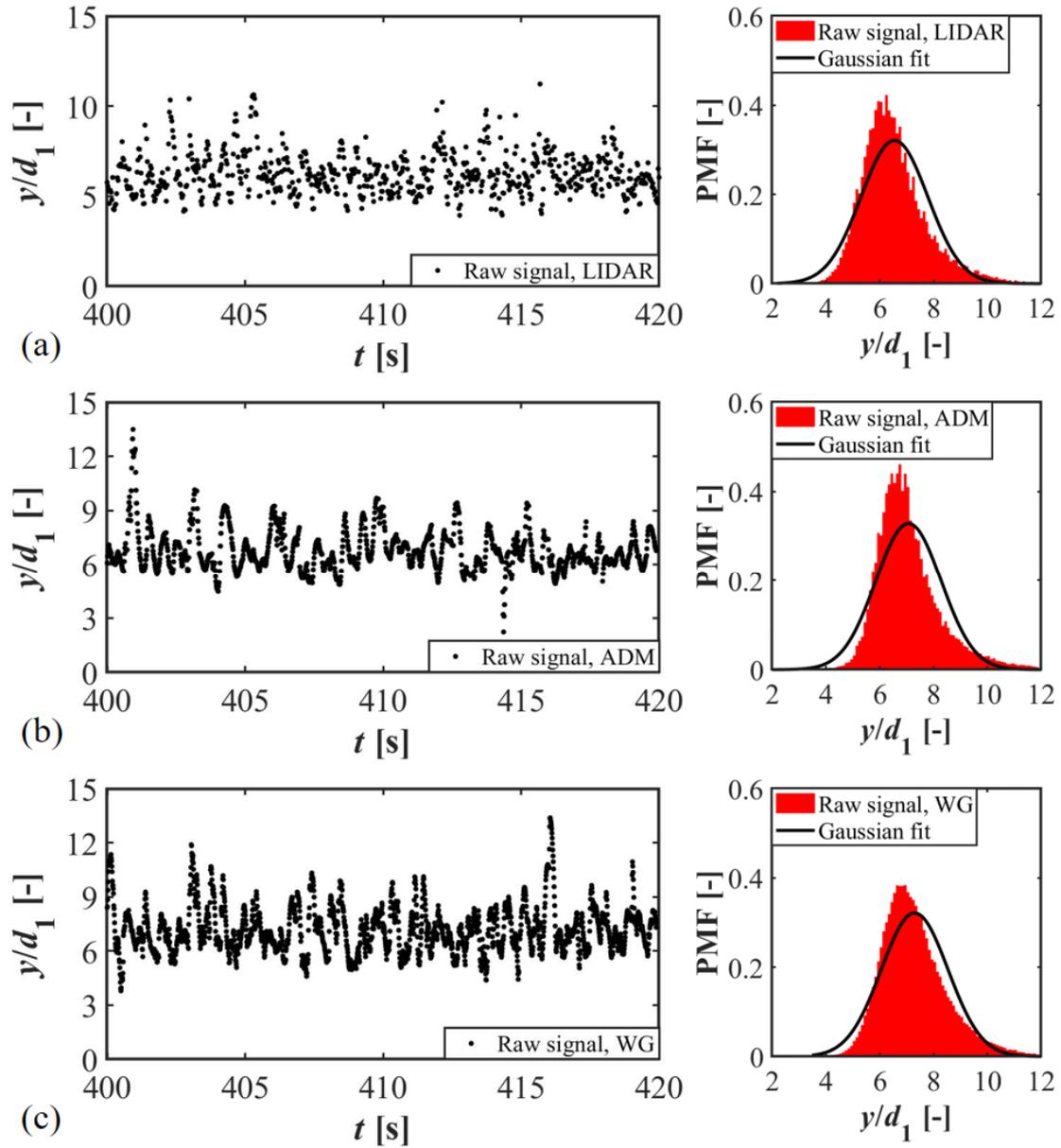

Figure 4. Selected raw data (left hand side) and PMFs for a signal of 1380 s (right hand side) at $(x-X_{toe,LIDAR})/L_r = 0.37$, $d_1 = 0.028$ m, $Fr_1 = 8$, $Re = 1.2 \times 10^5$: (a) LIDAR; (b) ADM; (c) WG.

Analysis of the raw data collected in the first two stages (see Section 2.1) provided basic free surface properties including the mean, standard deviation, skewness and kurtosis of the free surface elevations as well as the characteristic frequencies of the free surface motions. This is presented in Section 3 as well as in Supplementary Material S3. In addition, the LIDAR provided more detailed information about the time-varying jump toe position. Data collected during the third stage (see Section 2.1) was used to calculate the

auto- and cross-correlation functions of the free surface data and to estimate the free surface integral time $T_{xx}$ and length scales $L_{xy}$. These are presented in Section 4 and in Supplementary Material S4. The calculation of $L_{xy}$ was based upon the integration of the maximum cross-correlation coefficients $R_{xy,max}$ between two data points with distance $\Delta x$ up to the maximum distance $\Delta x_{max}$ [26,29]:

$$L_{xy} = \int_{x=0}^{x=\Delta x_{max}} R_{xy,max}(x) \, dx \qquad (1)$$

The integration was conducted using trapezoidal rule and different integration limits for $\Delta x_{max}$ were tested. For the continuous LIDAR data, $\Delta x_{max}$ was selected depending upon the distance to the first crossing of the $x$-axis, the distance with minimum $R_{xy,max}$ if no zero cross-correlation existed or the sensor distance complementing the maximum separation distance of ADMs and WGs respectively.

## 3. Understanding differences in basic free surface properties

This section systematically investigates differences in basic free surface properties of previous hydraulic jump studies using the simultaneously recorded data of three different instruments in the same flume and for the same flow conditions. While the full range of free surface properties was calculated for all flow conditions, only the key findings are presented here. Supplementary Material S3 includes additional results for skewness, kurtosis and characteristic frequencies.

This section starts with a comparative analysis of the free surface elevations and fluctuations using the sluice gate as the reference frame (Section 3.1). This allows the identification of potential instrumentation effects. In Section 3.2, the present data are compared with data from previous studies with comparable hydraulic jumps (Table 1) to assess potential differences with previous studies. This is conducted using $X_{toe,visual}$. Afterwards the differences in basic free surface properties derived from the present experiments with the same flow conditions are discussed using $X_{toe,visual}$ as the reference frame. Section 3.3 introduces a potential solution for the alignment of free surface elevations and fluctuations using $X_{toe,LIDAR}$. The findings show that human bias in determining $X_{toe,visual}$ can explain many of the large differences in previous studies, while the results obtained from different instruments are consistent when referenced to $X_{toe,LIDAR}$.

*3.1 Comparison of instrumentation shows close agreement*

Figure 5 shows representative mean free surface elevations and fluctuations for the three hydraulic jumps and for the three instruments. All sub-figures are shown in dimensional terms and relative to the sluice gate as a horizontal reference frame to provide important information of potential effects of instrumentation on basic free surface properties. Note that only data for one of the repeated experiments are shown, but that the comparison between the results for each experimental run provided the same overall findings.

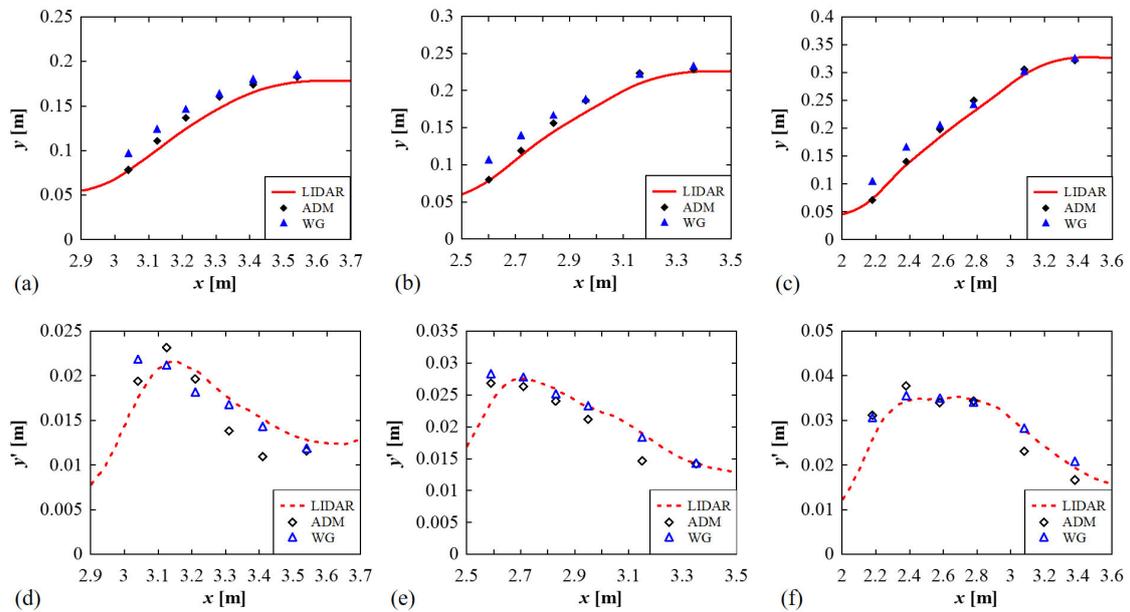

Figure 5. Mean free surface elevations (top row) and free surface fluctuations (bottom row) relative to the upstream sluice gate for one repeat: left column (a & d) $Fr_1 = 3.5$; $Re = 9.2 \times 10^4$; middle column (b & e) $Fr_1 = 5$; $Re = 1 \times 10^5$; right column (c & f) $Fr_1 = 8$; $Re = 1.2 \times 10^5$.

The comparison of the three instrumentations showed close agreement of the overall distribution and shape of both mean elevations (Figure 5, a-c) and free surface fluctuations (Figure 5, e-f), while distinct differences in magnitudes recorded with each instrument were observed. Visually, there was better agreement for higher Froude numbers. The elevations recorded with the ADMs (black symbols) were similar to the LIDAR data (red line) at the start of the hydraulic jump (first 1-2 comparison points) but were consistently above the LIDAR in the latter part of the roller irrespective of the flow conditions. Slightly higher mean free surfaces are most likely a result of the ADM spot size and the instrument recording the first return on a sloped surface [25,26,37]. Maximum differences in mean elevations between LIDAR and ADMs of 10% were observed in the centre of the roller, while the differences in all other flow regions were 5% on average.

The mean surface elevations measured by WGs (blue symbols) were consistently above the LIDAR and ADM data in the first half of the roller, while the differences decreased towards the end of the roller. The WGs measured the flow depth intrusively resulting in local bulking of water in front of the WG stem, which was most pronounced in regions of largest flow velocities. In addition, the wetting and drying processes of the WGs, as well as impacts of free surface splashing onto the wires may result in larger free surface elevations in the first part of the roller. The maximum differences in elevations between LIDAR and WG were observed at the first comparison point with differences of 23%, 30% and 35% for $Fr_1$ = 3.5, 5 and 8, respectively.

The free surface fluctuations $y'$ for WGs and LIDAR were in close agreement along the first half of the jump roller with average differences of 8% for all flow conditions (Figure 5 e-f). Average differences both increased (12% for $Fr_1$ = 3.5) and decreased (5% for $Fr_1$ = 5 and 8) for the second half of the roller. The comparison of ADM and LIDAR data showed an average 9% larger values of $y'/d_1$ for the ADMs for the first half of the roller, while the LIDAR free surface fluctuations were comparatively larger for the middle part of the roller with maximum differences of 32%, 25% and 18% for $Fr_1$ = 3.5, 5 and 8. In the latter portion of the roller, the relative differences were less than 14%.

In summary, the simultaneous measurements with each instrument showed comparable results, albeit some differences, for each individual experiment. Similarly, close agreements were found in terms of skewness, kurtosis and free surface frequencies (See Supplementary Material S3). This finding is significant since it suggests that (1) each instrument can be used for the recording of basic free surface properties and (2) suggests that previously reported large differences in free surface elevations and fluctuations may not be explained with the instrumentation but may be based upon other experimental effects.

*3.2 Comparative analysis of present and previous free surface properties using the visual mean jump toe*

This section further explores possible differences between various experimental results presented in the literature due to frame of reference. Specifically, the visual inspection of the mean jump toe position as the reference frame is common practice yet is highly subjective and prone to experimental bias. Figure 6 presents a dimensionless comparison of mean free surface profiles and fluctuations for 6 repeated experiments from the present study with identical inflow conditions using $X_{toe,visual}$ as the reference frame. In addition, data from previous experimental studies with similar flow conditions were reanalysed and included. Figure 6 shows large variations between repeated experiments from the present study with variations of up to 43% and 48% in $y$ and $y'$, respectively, while data from previous studies differed by up to 80% (even more just downstream of the jump toe). While the present experiments were carefully conducted with identical inflow conditions and the tail gate adjusted to achieve the visually same jump toe position, slight variability in the mean jump toe position occurred. This is due to the complexity of the hydraulic jump, its longitudinal oscillations, and non-uniform jump toe perimeter with an average convex shape [10,11,14]. Note that the differences in LIDAR data just downstream of the jump toe between present study and Montano et al. [11] was linked with the different approaches in replacing depth data below $d_1$ (see Section 2.1) and that the data also include the observer bias using $X_{toe,visual}$.

The free surface fluctuations $y'/d_1$ of all studies showed similar patterns along the jump roller. Just downstream of the jump toe, the free surface fluctuations increased sharply followed by a continuous decrease in $y'/d_1$ further downstream (Figure 6b). Large differences in $y'/d_1$ were consistently observed for repeated experiments in the present study as well as for previous studies (Figure 6b).

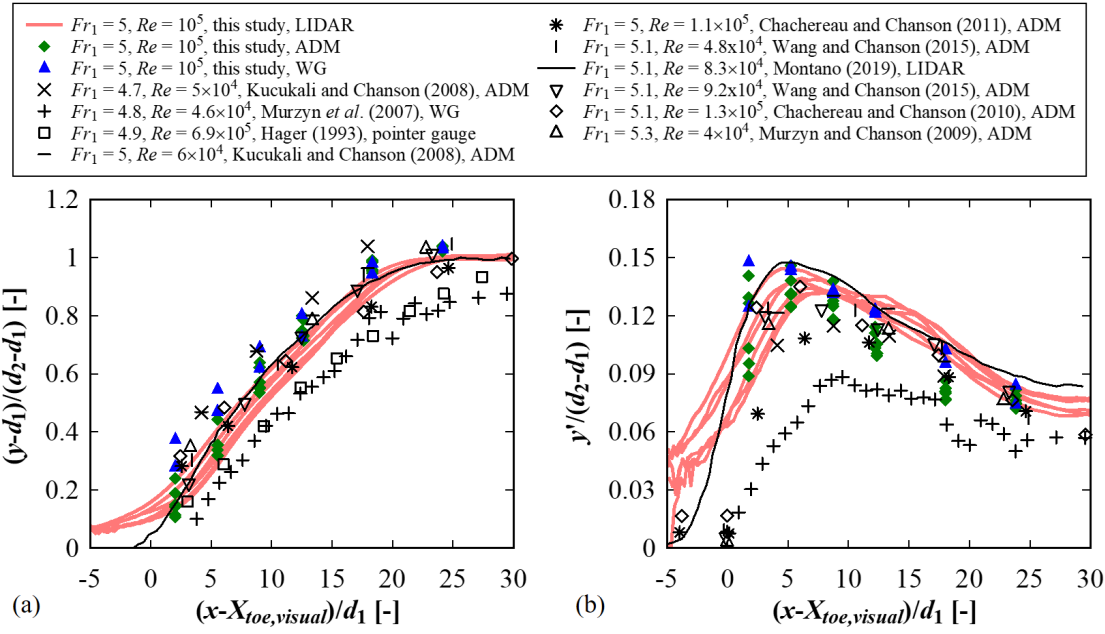

Figure 6. Comparison of present mean free surface elevations and free surface fluctuations with previous studies using LIDAR, ADM and WG relative to the visually observed mean jump toe: (a) mean free surface elevation; (b) free surface fluctuations.

While the comparative analysis in Figure 6 does not provide any information on the effects of different experimental facility, different inflow conditions, post-processing of experimental data and instrumentation, the data from the present study suggest a distinct effect of the visually determined jump toe location. In fact, shifting some of the previous data (as well as present studies) in the *x* direction could result in a closer agreement of free surface elevations and fluctuations. While this cannot be further explored for previous studies, the present data are further investigated and the effects of the observer bias in determining $X_{toe,visual}$ removed in the following section.

*3.3. Aligning free surface properties using the mean jump toe measured with LIDAR*

In the present study, the instantaneous jump toe was recorded by the LIDAR such that $X_{toe,LIDAR}$ was systematically calculated for all experiments. An example time series of the instantaneous jump toe positions $x_{toe}$ recorded with the LIDAR and its variation around the mean ($X_{toe,LIDAR}$) is shown in Figure 7. The figure shows small variations in the jump toe location including fast fluctuations with characteristic frequencies of 0.8 – 1 Hz for all flow conditions, as well as larger jump toe oscillations. Figure 7b-d show typical PMFs of $x_{toe}$ relative to $X_{toe,LIDAR}$. With increasing Froude numbers, the PMFs flattened and widened due to stronger jump toe motions with maximum amplitude of up to 0.6 m for

$Fr_1 = 8$. Figure 7 further emphasizes potential difficulties in visually determining the mean jump toe position which becomes harder with increasing $Fr_1$.

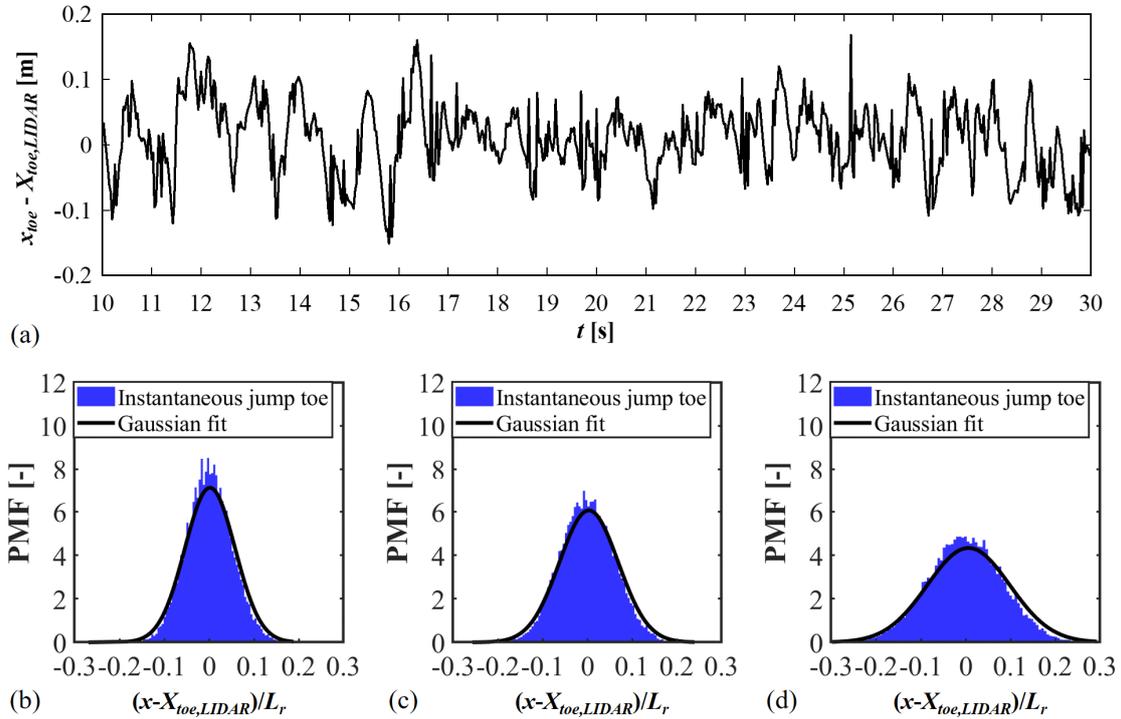

Figure 7. Segment (upper plot) and PMFs (lower plots) of the instantaneous jump toe positions recorded with the LIDAR: (a) $Fr_1 = 5$; (b) $Fr_1 = 3.5$; (c) $Fr_1 = 5$; (d) $Fr_1 = 8$.

To better quantify potential differences between $X_{toe,visual}$ and $X_{toe,LIDAR}$, the mean jump toe positions for both methods were systematically observed and the differences calculated as $X_{toe,LIDAR} - X_{toe,visual}$ for each of the experiments (Figure 8). This provides important guidance on the accuracy of visual mean jump toe positioning. While the median of $X_{toe,LIDAR} - X_{toe,visual}$ was close to 0 for all flow conditions, the visual observations may have differed from the LIDAR data by up to 0.055 m, 0.06 m and 0.07 m for $Fr_1 = 3.5$, 5 and 8, respectively. Such differences are significant considering the rapid longitudinal motions of the hydraulic jump (Figure 1b) and that the roller length of hydraulic jumps in laboratory conditions is often in the range of 0.3 to 1.4 m.

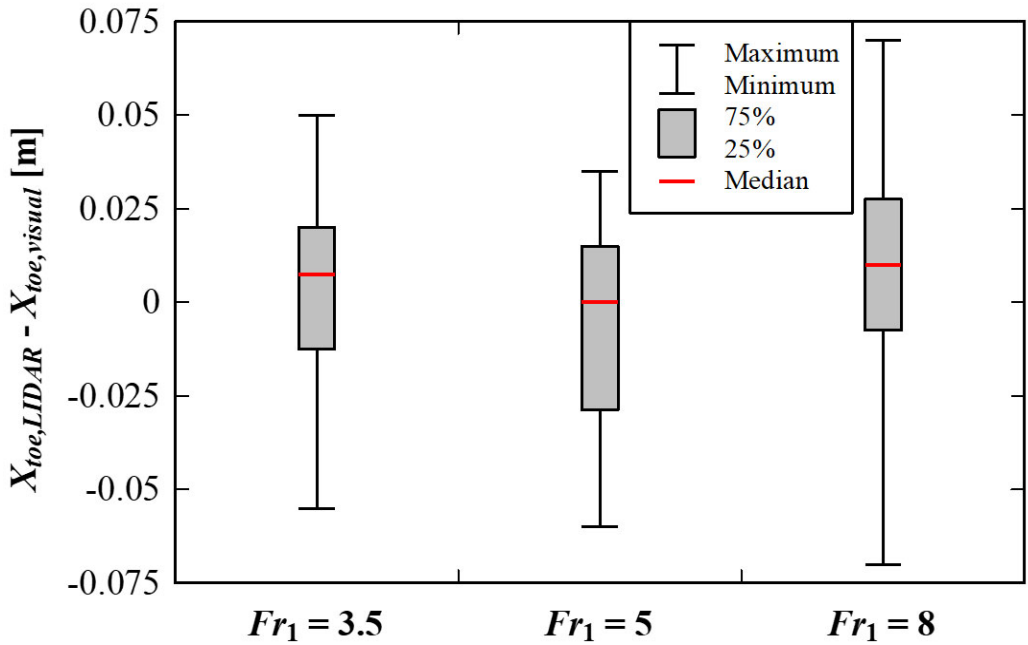

Figure 8. Differences between visual observations and LIDAR measurements of the mean jump toe position; Box and Whisker plot of all present experiments (12 runs for $Fr_1$ = 3.5; 12 runs for $Fr_1$ = 5; 14 runs for $Fr_1$ = 8).

Figure 9 shows mean free surface profiles and free surface fluctuations using $X_{toe,LIDAR}$ as the reference frame for all instruments. The mean LIDAR data are shown as a continuous line including error bars representing the 5$^{th}$ and 95$^{th}$ percentiles from the repeat experiments, while the ADM and WG data of all repeated runs are shown as symbols as per their fixed measurement locations. Clearly better agreement between experiments is noticeable when using this new frame of reference (Figure 6 vs Figure 9a). All data now show close agreement in respective properties for the LIDAR, ADMs and WGs. The largest difference between 5$^{th}$ and 95$^{th}$ percentiles of all LIDAR data was observed close to the jump toe (($x$-$X_{toe,LIDAR}$)/$L_r$ ~ 0) with $y/d_1$ = 0.03 and $y'/d_1$ = 0.03 irrespective of the flow conditions. These differences included the effects of experimental repeatability, measurements at different longitudinal cross-sections with $\Delta z$ = ±70 mm, as well as potential flow disturbances due to the intrusive wire gauges. The present findings highlight that an alignment using $X_{toe,LIDAR}$ is much more consistent than using visual observations with $X_{toe,visual}$. This finding suggests that the bias/error induced from visual observation of the mean jump toe can explain some differences in previously reported studies. It is therefore recommended that any future measurements of flow properties in hydraulic jumps should simultaneously measure the instantaneous hydraulic jump toe position. As shown in Figure 9, it appears that LIDAR is a suitable instrument for this allowing the remote recording of $x_{toe}$ and $X_{toe,LIDAR}$, respectively, while the full range of free surface properties can be simultaneously and accurately recorded with high spatial and temporal resolution.

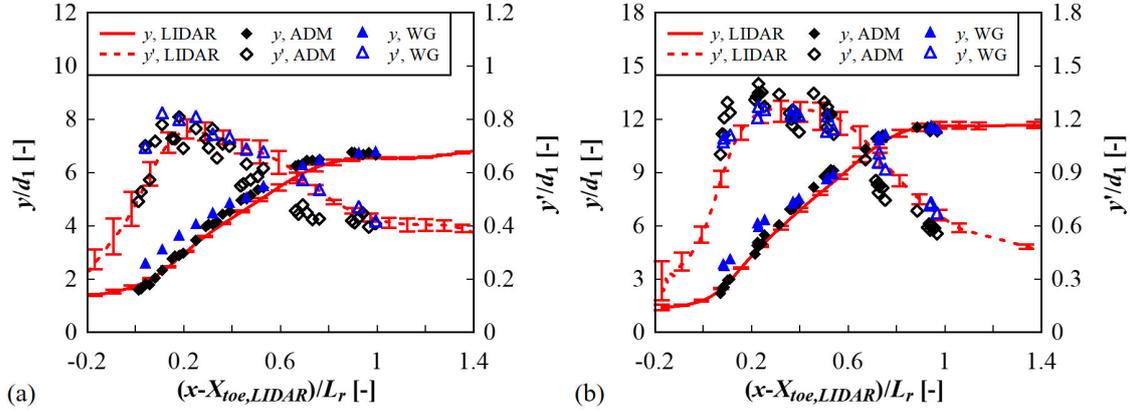

Figure 9. Mean free surface elevations and free surface fluctuations in aerated hydraulic jumps measured with LIDAR, ADM and WG: (a) $d_1$ = 0.034 m, $Fr_1$ = 5, $Re$ = $10^5$; (b) $d_1$ = 0.028 m, $Fr_1$ = 8, $Re$ = $1.2 \times 10^5$.

## 4. Assessing differences in free surface integral length scales

To assess the effects of instrumentation and post-processing methods on the free surface time and length scales, detailed measurements were conducted for all three types of instruments as part of the third stage of experiments (details in Section 2.1). Results of the auto-correlation analyses and the associated auto-correlation integral time scales are presented in Supplementary Material S4, while the results of the cross-correlation analysis and free surface length scales are shown below. Effects of high pass filtering and the integration length are also discussed.

Irrespective of the instrument, cross-correlation analysis was performed for simultaneously sampled consecutive data points separated by distance $\Delta x$. As the LIDAR records a continuous free surface, cross-correlations can be done at a large range of points, while the ADM and WG data are limited to a fixed set of discrete locations and this impacts the integrated length scales reported. Representative cross-correlation functions $R_{xy}$ are shown in Figure 10 comprising raw data (top row) and high pass filtered data (0.1 Hz) (bottom row) following the procedure of Chachereau and Chanson [42] to remove the slow jump toe motions from the raw signal. Overall, the shapes of the cross-correlation functions at respective locations were consistent between different instruments indicating that all instruments detected similar free surface features. Close agreement was observed for LIDAR and WG for all locations and Froude numbers (Figure 10, red and blue symbols). For the raw signals, ADM data (black symbols) showed larger $R_{xy}$, as well as a slight time-lag in the location of the peak $R_{xy}$ for low Froude numbers compared to WG and LIDAR but was in better agreement with LIDAR and WG for high Froude numbers (Figure 10a,d vs. Figure 10c,f).

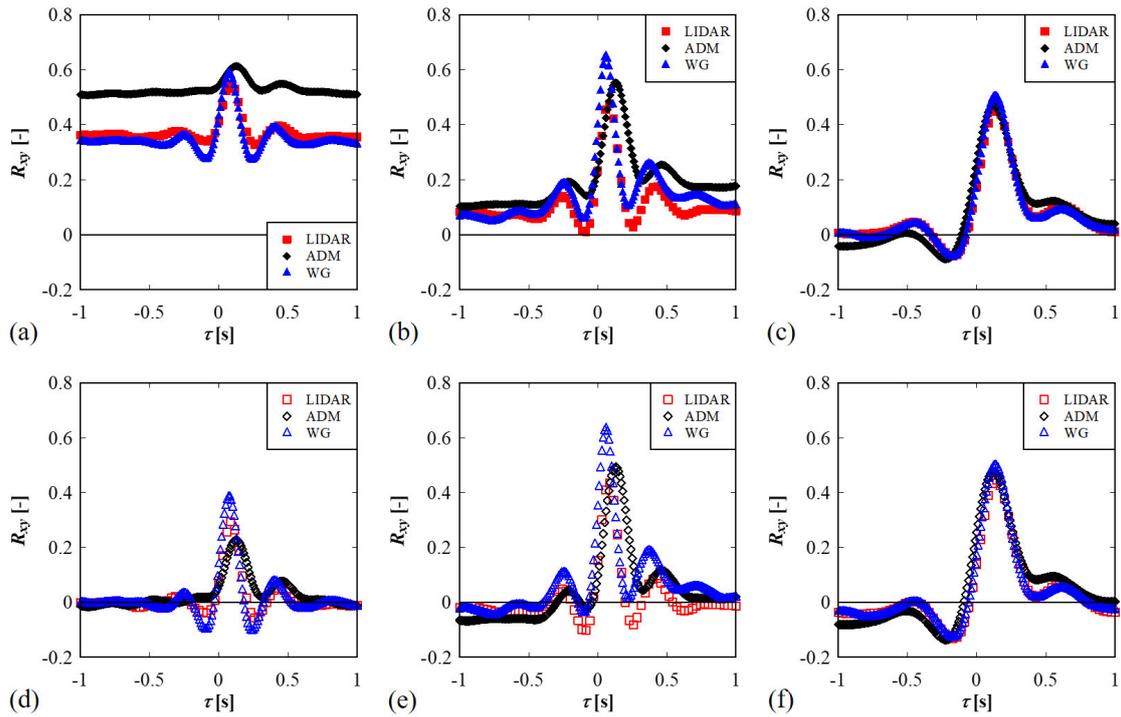

Figure 10. Cross-correlation functions of raw free surface data (top row) and high pass (0.1 Hz) filtered free surface data (bottom row) sampled at two locations separated by a distance $\Delta x$ measured with LIDAR, ADM and WG: (a & d) $Fr_1 = 3.5$, $(x-X_{toe,LIDAR})/L_r = 0.25$, $\Delta x/L_r = 0.15$; (b & e) $Fr_1 = 3.5$, $(x-X_{toe,LIDAR})/L_r = 0.68$, $\Delta x/L_r = 0.15$; (c & f) $Fr_1 = 8$, $(x-X_{toe,LIDAR})/L_r = 0.74$, $\Delta x/L_r = 0.14$

For the raw signals, the cross-correlation functions did not always cross the *x* axis for all instruments in particular in the region closest to the jump toe (Figure 10a) which was linked with the effects of the low frequency motions of the hydraulic jump. The high pass filtering resulted in a downwards shift in the cross-correlation functions and a crossing of the *x* axis ($R_{xy} = 0$) for all data and was therefore used for the calculation of the integral length scales presented here.

Equation (1) was applied to calculate $L_{xy}$ for all instruments and for various integration limits of $\Delta x_{max}$. Figure 11 shows dimensionless length scales $L_{xy}/d_1$ of the high pass filtered signals including a comparison of LIDAR with ADM for the present study with the maximum measurement range of ADM ($\Delta x_{max}$ = 285 mm) (Figure 11a) and LIDAR and WG data ($\Delta x_{max}$ = 475 mm) for the present study (Figure 11b). For completeness, a comparison with previous ADM data [26] (Figure 11c) and WG data [24] (Figure 11d) using the respective integration length $\Delta x_{max}$ of these studies are also presented. Note that the closest values of $\Delta x_{max}$ for the present ADMs and WGs was used for inter-study comparison. For the present study, the length scales of the LIDAR were calculated starting from a location downstream of $X_{toe,LIDAR}$ where less than 5% of the data was NaN and error bars are added in Figure 11 for the repeated tests. Note that Figure 11c and d used $X_{toe,visual}$ for previous ADM and WG data while all LIDAR data in Figure 11 used $X_{toe,LIDAR}$.

The comparative analysis revealed the following key findings. For the LIDAR data, an increase in the integration range $\Delta x_{max}$ resulted in an increase of the length scales which was consistent with the observations of Montano and Felder [29]. For all instruments used in this study, higher Froude numbers resulted in larger dimensionless length scales, while the dimensional length scales were similar irrespective of $Fr_1$. Encouragingly, the comparison of free-surface length scales between different instruments showed a relatively close agreement for the present study, particularly between LIDAR and WG (Figure 11b) albeit some data scatter was observed for the point source instruments that could be explained with the lower spatial resolution but may also indicate small instrument effects.

The comparative analysis with present data for similar $\Delta x_{max}$ showed similar trend and magnitude in length scales derived from ADM and LIDAR data with the dimensionless length scales of Chachereau and Chanson [26], while length scales based upon raw data would be significantly larger. The comparison of the WG data with those reported by Murzyn et al. [24] showed large differences despite similar integration length. It appears that the WGs or the experimental setup used by Murzyn et al. [24] might be different to the present study. This is not just evident in Figure 11d, but also in the comparison of basic free surface properties (Figure 6) which showed consistently lower free surface elevations and fluctuations for the data of [24] compared to experimental studies with similar flow conditions.

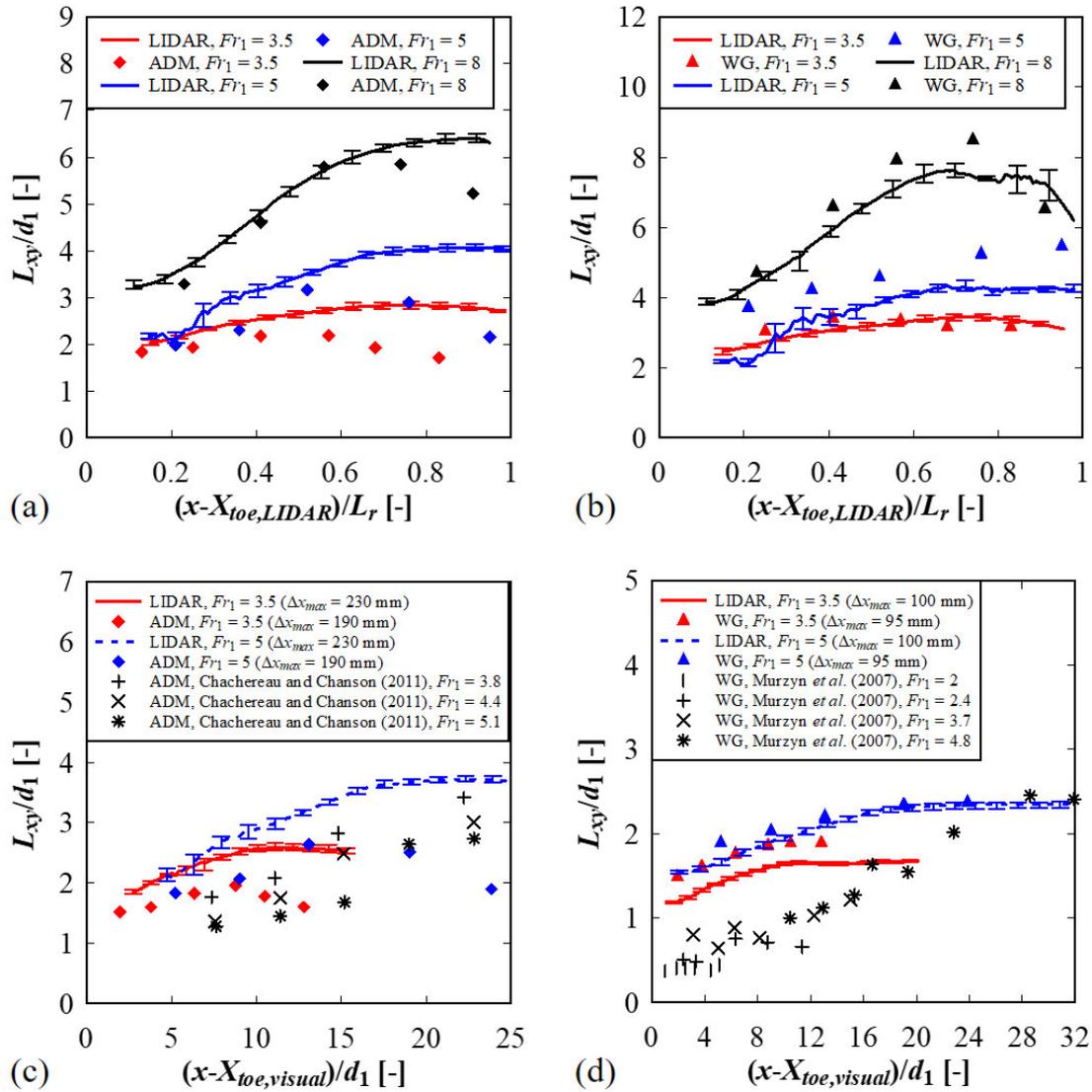

Figure 11. Free-surface integral length scales measured with LIDAR, ADM and WG for high pass filtered signals (0.1 Hz) with various integration lengths: (a) LIDAR and ADM data with $\Delta x_{max}$ = 285 mm; (b) LIDAR and WG data with $\Delta x_{max}$ = 475 mm; (c) comparison of present data with ADM data of Chachereau and Chanson 2011 ([26]) with $\Delta x_{max}$ = 230 mm, (d) comparison of present data with WG data of Murzyn et al. 2007 ([24]) with $\Delta x_{max}$ = 100 mm.

Overall, the agreement of LIDAR, WG and ADM data in the present study (Figure 11a,b) indicate that similar length scales can be measured with any of these instruments when the same post-processing and integration limits are applied. However, large differences can result from variations in the integration limit (Figure 11). Point source instruments are only able to record the data within the instrumentation space, while LIDAR data offers the advantage that length scales can be integrated to various length including to the crossing of the *x* axis. Since the size of the length scales is dependent on the integration length, future studies must report such properties as well as any post-processing of data thoroughly.

## 5. Conclusion

Free surface properties simultaneously measured with ADMs, WGs and LIDAR were for the first time compared in aerated hydraulic jumps with Froude numbers between 3.5 and 8.0. A summary of the key findings of this comparison is provided in Table . Strong similarity between instrumentations was observed for individual repeats of all flow conditions. However, strong jump toe movements resulted in imprecise visual observations of the mean jump toe location, which caused large variations between repeats using the sluice gate as a static reference frame. With the adjustment relative to the mean jump toe location recorded with the LIDAR, all three instruments recorded similar basic free surface properties. Small differences in magnitudes were observed in basic free surface properties including mean free surface elevations, standard deviations, and characteristic frequencies, while some larger differences were observed in terms of free surface time and length scales. The differences were smaller when data was high pass filtered, while differences with previous studies are still unresolved, but integration lengths and accurate measurement of $X_{toe}$ are likely to be contributing factors. Table 2 summarises the results of the comparative analysis between LIDAR, ADMs and WGs respectively.

Table 2. Summary of comparative analysis of LIDAR, WGs and ADMs in fully aerated hydraulic jumps with $3.5 \leq Fr_1 \leq 8$, $9.2 \times 10^4 \leq Re \leq 1.2 \times 10^5$.

| Free-surface parameter | Comparative analysis between | |
|---|---|---|
| | LIDAR and ADMs | LIDAR and WGs |
| Mean elevation | ADMs larger (6% on average) irrespective of flow conditions | WGs larger (13% on average) irrespective of flow conditions |
| Standard deviation | ADM larger (smaller) for $(x-X_{toe,LIDAR})/L_r <(>) 0.4$ (12% on average); difference decreases with $Fr_1$ | WG larger (smaller) for $(x-X_{toe,LIDAR})/L_r <(>) 0.4$ (10% on average); difference decreases with $Fr_1$ |
| Skewness | rsmd = 0.28; difference decreases with $Fr_1$ | rmsd = 0.2 irrespective of flow conditions |
| Kurtosis | rsmd = 1.2; difference decreases with $Fr_1$ | rsmd = 0.8; difference increases with $Fr_1$ |
| Characteristic frequency | Similar irrespective of flow conditions; frequency peaks for ADMs less pronounced | Similar irrespective of flow conditions |
| Auto-correlation integral time scale (High-pass signal) | ADMs larger (42% on average) irrespective of flow conditions | Small difference (7% on average) irrespective of flow conditions |
| Cross-correlation integral length scale (High-pass signal) | LIDAR larger (19% on average); difference decreases with $Fr_1$ | WGs larger (11% on average) irrespective of flow conditions |

Overall, LIDAR, ADM and WG provided similar distribution patterns in all investigated free surface properties. This finding is important since it suggests that any of these instruments can be used for the detection of free surface properties in hydraulic jumps and variations between studies may be a result of frame of reference. Importantly, LIDAR demonstrated the ability to simultaneously track the jump toe positions, which enables the alignment of free surface properties to the true mean jump toe position, providing more consistent measurement results. It is suggested that future studies of any flow properties of hydraulic jumps should simultaneously measure the time-varying jump toe location to report more accurate and consistent results relative to the mean jump toe. More generally other open channel flow phenomena with strong time varying properties such as breaking waves and tidal bores or with strong spatial differences such as shock waves, standing waves or jets should consider instrumentation that allows the recording of free surface properties with high spatial and temporal resolution rather than to rely on point measurements.

**Acknowledgements**

The authors thank Rob Jenkins (Water Research Laboratory, UNSW Sydney) for the technical assistance. They thank Dr Laura Montano (Water Research Laboratory, UNSW Sydney) and Dr Matthias Kramer (UNSW Canberra) for fruitful discussions.

**List of Symbols**

| | |
|---|---|
| $d_1$ | inflow depth [m] |
| $d_2$ | downstream conjugate depth [m] |
| $F_{fs}$ | characteristic free surface frequency [Hz] |
| $Fr_1$ | inflow Froude number [-] |
| $g$ | gravity acceleration constant [m$^2$/s] |
| $L_r$ | roller length of the hydraulic jump [m] |
| $L_{xy}$ | free surface cross-correlation integral length scale [m] |
| $q$ | discharge per unit width [m$^2$/s] |
| $Re$ | Reynolds number [-] |
| $R_{xx}$ | auto-correlation function [-] |
| $R_{xx,min}$ | minimum auto-correlation coefficient [-] |
| $R_{xy}$ | cross-correlation function [-] |
| $R_{xy,max}$ | maximum cross-correlation coefficient [-] |
| $R_{xy,min}$ | minimum cross-correlation coefficient [-] |
| $R_{xz}$ | cross-correlation function between different sensors [-] |
| $T_{xx}$ | free surface auto-correlation integral time scale [s] |
| $t$ | time [s] |
| $v_1$ | depth average inflow velocity [m/s] |
| $W$ | width of the channel [m] |
| $X_{toe}$ | mean jump toe position [m] |
| $X_{toe,LIDAR}$ | mean jump toe position measured with the LIDAR [m] |
| $X_{toe,visual}$ | visually determined mean jump toe position [m] |
| $x$ | longitudinal distance relative to the mean jump toe position [m] |
| $x_{toe}$ | instantaneous jump toe position measured with the LIDAR [m] |
| $y$ | vertical distance above the channel bed [m] |
| $z$ | transverse distance relative to the centerline of the channel [m] |
| $\Delta X_{ADM}$ | longitudinal distance between consecutive ADMs [m] |
| $\Delta x$ | distance between sampling points [m] |
| $\Delta x_{max}$ | maximum integration distance for free surface integral length scales [m] |
| $\Delta z$ | transverse distance between instruments [m] |
| $v$ | water kinematic viscosity [m$^2$/s] |

Supplementary Materials to manuscript

"Aligning free surface properties in time-varying hydraulic jumps"

Rui LI[1], Kristen D. SPLINTER[2] and Stefan FELDER[3]

[1] PhD Candidate, UNSW Sydney, Water Research Laboratory, School of Civil and Environmental Engineering, 110 King St, Manly Vale, NSW, 2093, Australia
ORCID: 0000-0002-7387-8004

[2] Senior Lecturer, UNSW Sydney, Water Research Laboratory, School of Civil and Environmental Engineering, 110 King St, Manly Vale, NSW, 2093, Australia
ORCID: 0000-0002-0082-8444

[3] Senior Lecturer, UNSW Sydney, Water Research Laboratory, School of Civil and Environmental Engineering, 110 King St, Manly Vale, NSW, 2093, Australia
+61 (2) 8071 9861; s.felder@unsw.edu.au (corresponding author)
ORCID: 0000-0003-1079-6658

**Introduction**

This supplementary material provides complementary information for the main manuscript. Section S1 discusses the effects of transverse sampling on the reported basic free surface properties. S2 discusses the effects of different filter methods on the free surface properties. In Section S3, the results of additional free surface properties are presented including skewness, kurtosis and characteristic frequencies that complement the free surface properties of mean profiles and fluctuations presented in the main manuscript. Section S4 presents advanced free surface properties including the free surface auto-correlation functions and the free surface time scales.

**S1. Effects of transverse instrument separation on free surface properties**

As detailed in Section 2.1 (main manuscript) in order to simultaneously collect data with the three instrumentations, LIDAR, WGs and ADMs were separated by a transverse distance $\Delta z$ = 0.07 m between LIDAR and ADM/WG or $\Delta z$ = 0.014 m between ADM and WG (see Figure 2, main manuscript). To determine if this small transverse spacing influenced the reported results, a series of experiments were conducted whereby the LIDAR and ADM instruments were rotated between each of the three transverse locations and mean free surface properties compared. As shown in Figure S1 for the 5 repeat LIDAR tests, mean and free surface fluctuations when referenced to the mean jump toe derived from the LIDAR $X_{toe,LIDAR}$ are almost identical, indicating that transverse variability in bulk statistical properties of the hydraulic jumps are minimal. Similarly, the 2 repeat ADM tests show very good agreement with small longitudinal differences due to the fixed-point source measurements and the variations in the mean jump toe between different experiments. These tests also reaffirm that differences in the reported basic free surface properties are due to differences in instrument signal capture (e.g. ADM vs. LIDAR) and not due to the small transverse offsets between instruments.

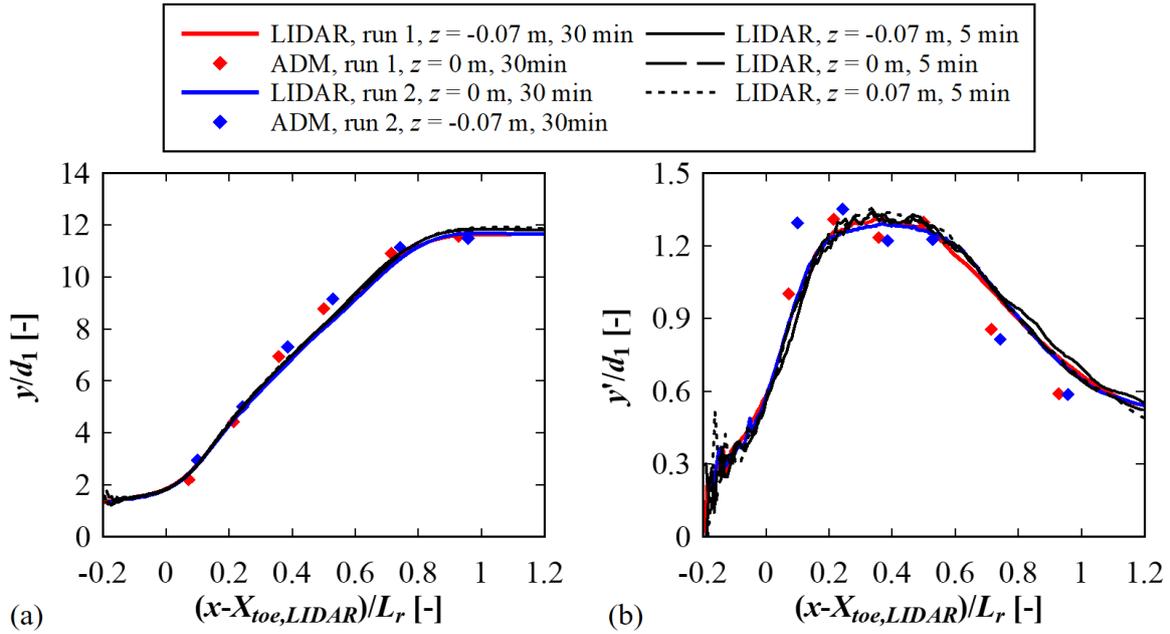

Figure S1. Comparison of (a) mean free surface elevations and (b) fluctuations recorded by LIDAR (lines) and ADM (symbols) at $z$ = -0.07, 0, 0.07 m: $Fr_1$ = 8; $Re$ = 1.2×10$^5$.

Cross-correlation analysis between different instruments were also done to test the agreement in instantaneous data between different instruments. Example cross-correlation functions $R_{xz}$ for the hydraulic jump with $Fr_1$ = 5 (Figure S2) and $Fr_1$ = 8 (Figure S3) are provided. Results for $Fr_1$ = 3.5 were very similar to $Fr_1$ = 5. LIDAR along the centreline was used as the first sensor in the cross-correlation analysis between LIDAR and ADM ($z$ = -0.07 m) or WG ($z$ = 0.07 m), and ADM was the first sensor to calculate $R_{xz}$ between ADM and WG. A positive lag indicates that the second sensor leads the first.

The cross-correlation analysis shows a strong correlation between all sensors despite the transverse distance. $R_{xz}$ is consistently lower for ADM and WG due to the larger separation distance. The results show that small transverse spacing of the instruments was unlikely to significantly influence the free surface properties reported in the manuscript.

A small time lag $\tau$ was consistently observed between LIDAR and ADM, and WG and ADM (Figure and Figure S3, respectively) indicating that the ADM recorded a slightly earlier signal. Larger lags were observed for larger $Fr_1$. This is most likely due to the larger spot size of the ADM and the instrument recording the first return signal on the sloping roller. Maximum cross-correlations between LIDAR and WG were consistently close to $\tau$ = 0 for all inflow conditions, indicating that both instruments recorded very similar time-dependent properties.

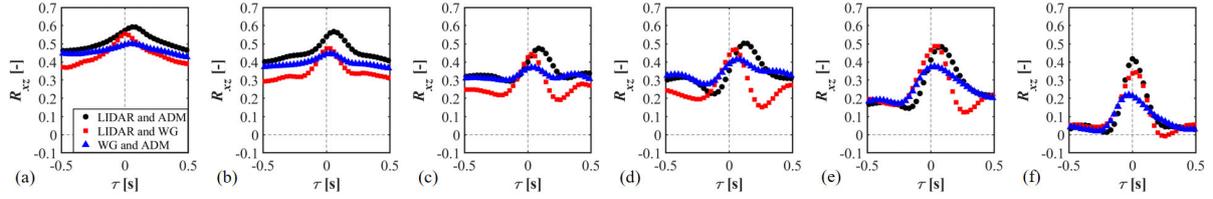

Figure S2. Transverse cross-correlation between LIDAR ($z = 0$), ADM ($z = -0.07$ m) and WG ($z = 0.07$ m) for $Fr_1 = 5$ and $Re = 10^5$ using raw signals: (a) $(x-X_{toe,LIDAR})/d_1 = 0.04$; (b) $(x-X_{toe,LIDAR})/d_1 = 0.18$; (c) $(x-X_{toe,LIDAR})/d_1 = 0.32$; (d) $(x-X_{toe,LIDAR})/d_1 = 0.46$; (e) $(x-X_{toe,LIDAR})/d_1 = 0.69$; (f) $(x-X_{toe,LIDAR})/d_1 = 0.92$.

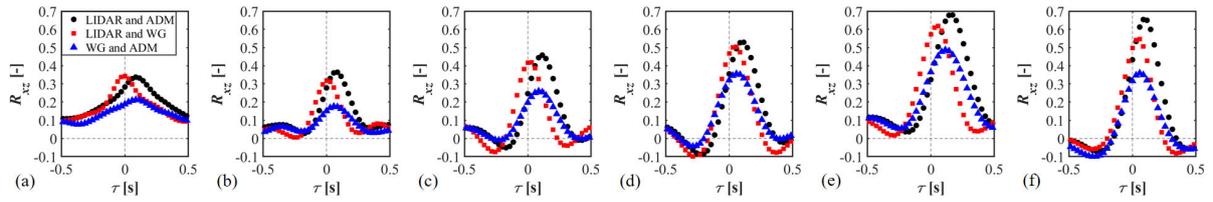

Figure S3. Transverse cross-correlation between LIDAR ($z = 0$), ADM ($z = -0.07$ m) and WG ($z = 0.07$ m) for $Fr_1 = 8$ and $Re = 1.2 \times 10^5$ using raw signals: (a) $(x-X_{toe,LIDAR})/d_1 = 0.09$; (b) $(x-X_{toe,LIDAR})/d_1 = 0.23$; (c) $(x-X_{toe,LIDAR})/d_1 = 0.37$; (d) $(x-X_{toe,LIDAR})/d_1 = 0.51$; (e) $(x-X_{toe,LIDAR})/d_1 = 0.73$; (f) $(x-X_{toe,LIDAR})/d_1 = 0.94$.

## S2. Effects of signal filtering on free surface properties

*Removal of outlier data*

The raw signals of ADMs and WGs revealed very few spikes (see Figure 4 in manuscript) and no obvious signal spikes had to be removed. However, different filtering methods were tested on the present data to determine if filtering impacted on the basic and advanced free surface properties reported due to the removal of outlier and signal de-spiking [1,2].

Outlier removal was tested using a threshold of $\lambda = \sqrt{2 \ln(N)}$ times the standard deviation, where $N$ is the number of sampled points [1,2]. For the present experiments $\lambda > 4.46$ resulting in filtered data of less than 0.5% along the hydraulic jump. Signal de-spiking using an elliptical bound on both recorded water depth and its derivative (vertical velocity) was also tested [1,2]. This filter technique was repeated until no more data was removed. Overall, less than 1.4% of data was filtered along the hydraulic jump using this method. An example of the filtered signal for the strongest hydraulic jump with $Fr_1 = 8$ in the most violent region at $(x-X_{toe,\,LIDAR})/L_r = 0.37$) is presented in Figure S4. As visible in Figure S4, the filtered data did not necessarily represent erroneous data or outliers. Therefore, to not filter potentially valid data, no filtering was applied to any instrumentation along the jump roller.

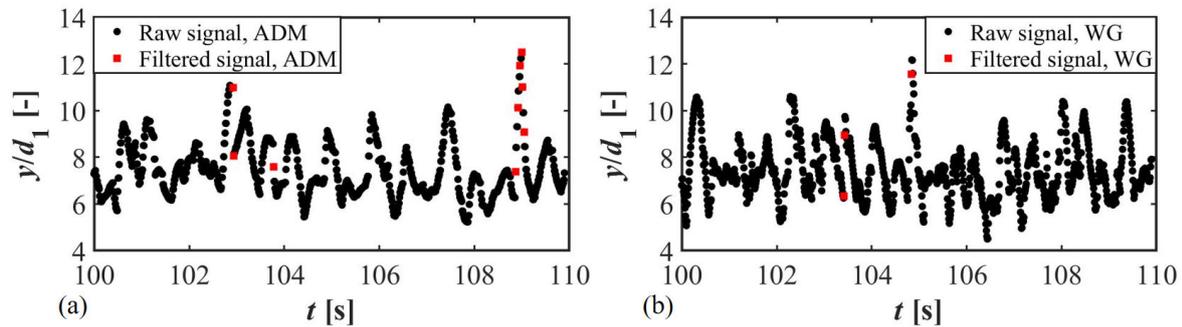

Figure S4. Examples of raw and filtered signal using de-spiking technique (Valero *et al.* 2020) for $Fr_1 = 8$ and $Re = 1.2 \times 10^5$ at $(x-X_{toe,LIDAR})/L_r = 0.37$: (a) ADM sensor; (b) WG sensor.

Effects of signal filtering on the free surface properties were tested including mean, standard deviations, frequencies, as well as time and length scales. Figure S5 compares free-surface properties calculated for the raw and filtered signals (elliptical bound filter [2]). The maximum difference in free-surface mean $y$ and standard deviations $y'$ between raw and filtered signal was less than 1.7 mm (3.3%) and 1.5 mm (18%) at the locations close to the jump toe, while they were much lower further downstream. For the strongest hydraulic jump with $Fr_1 = 8$, raw and filtered signals for all instruments had a root mean square differences (rmsd) of less than 0.38 and 2.6 in skewness and kurtosis, respectively. The strong effect of filtering on skewness (Figure S5c) and kurtosis (Figure S5d) close to the mean jump toe location was related with jump toe motions, which resulted in a long positive tail in the signal distributions. These oscillations were meaningful data, as such, no filtering was applied. Filtering did not lead to any significant differences in the FFT as well as the auto- and cross-correlation functions since any filtered data were replaced by interpolated data. Therefore, negligible differences were observed in characteristic frequencies and free-surface time and length scales.

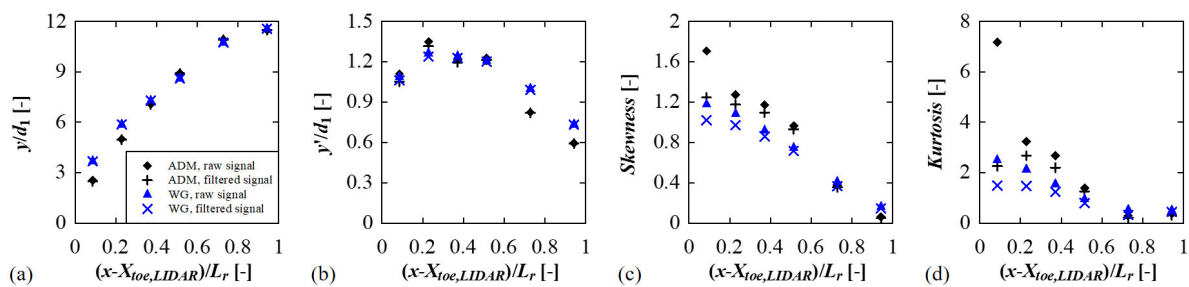

Figure S5. Free-surface properties along the roller for $Fr_1 = 8$ and $Re = 1.2 \times 10^5$ calculated using raw signal and filtered signal of ADMs and WGs: (a) mean free-surface profile; (b) free-surface fluctuations; (c) Skewness; (d) Kurtosis.

Filtering approaches for LIDAR were tested by Li *et al*. [3]. Montano *et al*. [4] and Li *et al*. [3] identified outliers using both time and space filtering. However, they showed that such filtering may remove valid data points in the roller region and filtering was therefore only applied downstream of the roller. Downstream of $L_r$, LIDAR data were filtered using 3 standard deviations of 12 neighbourhood points in the space domain and 4 standard deviations of 12 neighbourhood points in the time domain [3].

*Effect of instrument sampling resolution*

LIDAR signals were recorded with 35 Hz while WGs and ADMs were acquired with 100 Hz. To test potential effects of different sampling frequencies, the original sampling frequency of ADMs and WGs were down sampled to 33.3 Hz, to closely resemble the 35 Hz sampling frequency of the LIDAR. Minimal differences in free surface properties were identified between raw data and down sampled signals in terms of all free surfaces properties. As an example, the results of a Fast Fourier Transform (FFT) for both raw and down-sampled signals are shown in Figure S6. The results indicate little differences in terms of the dominant peak, while the decay of the raw signal (Figure S6a) was slightly steeper after the peak.

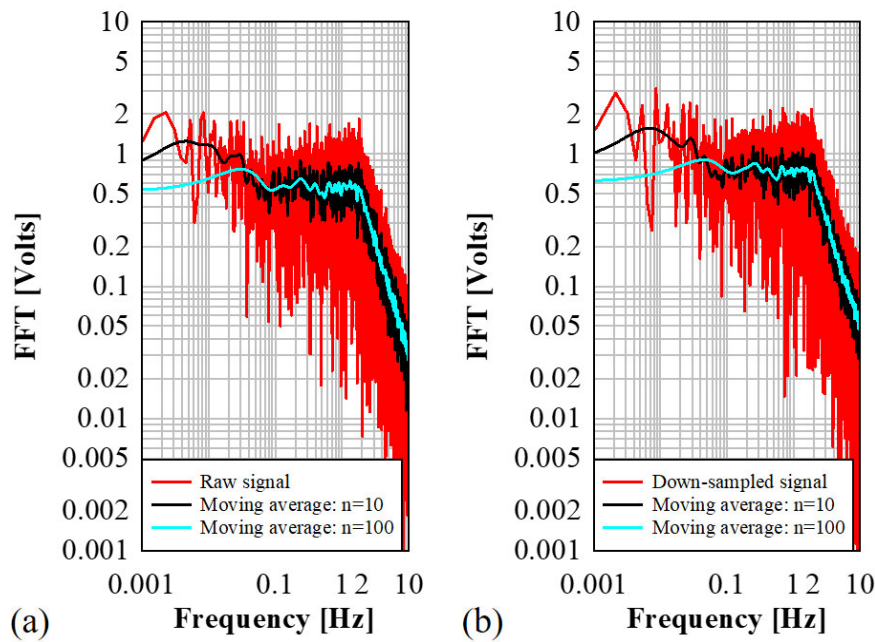

Figure S6. FFT analysis of an ADM signal for $Fr_1 = 8$ at $(x-X_{toe,LIDAR})/L_r = 0.39$ using (a) raw signal; (b) down-sampled signal.

## S3. Basic free-surface properties: skewness, kurtosis and characteristic frequencies

The alignment of mean free surface elevations and free surface fluctuations using $X_{toe,LIDAR}$ was discussed in Section 3 of the manuscript. Further basic free surface properties including skewness, kurtosis and characteristic frequencies were investigated using $X_{toe,LIDAR}$ as the reference frame for the three instrumentations. Example skewness and kurtosis distributions are shown in Figure S7. Overall, there was good agreement between the results derived from each of the individual instruments. With increasing Froude numbers, the overall magnitude of skewness increased irrespective of the instrumentation. This is a direct result of more water ejections and splashes resulting in a skewness towards higher recorded free surface elevations for higher Froude numbers. The skewness distributions decreased along the jump roller for all instruments and this trend was steeper for higher Froude numbers. This decreasing trend is a result of less ejections and splashing away from the jump toe. The data distribution at the end of the roller (($x-X_{toe,LIDAR}$)/$L_r$ =1) was close to normal with a skewness of 0 irrespective of flow conditions. For $Fr_1$ = 8, both ADMs and WGs showed a rms difference (rmsd) of 0.24 compared to LIDAR. For lower Froude numbers, there was better agreement between WG and LIDAR (rmsd = 0.19) compared to ADM and LIDAR (rmsd = 0.3).

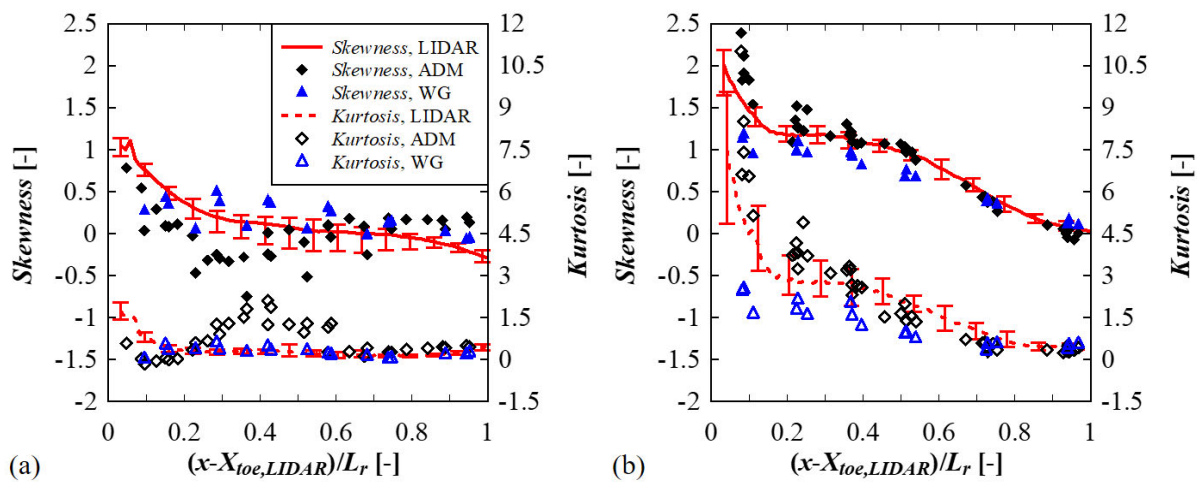

Figure S7. Skewness and excess kurtosis of the free surface in aerated hydraulic jumps measured with LIDAR, ADM and WG (error bars indicate the 5$^{th}$ and 95$^{th}$ percentiles of repeated runs for LIDAR): (a) $d_1$ = 0.041 m, $Fr_1$ = 3.5, $Re$ = 9.2×10$^4$; (b) $d_1$ = 0.028 m, $Fr_1$ = 8, $Re$ = 1.2×10$^5$.

In examining excess kurtosis (i.e. kurtosis – 3) to determine the heaviness of the tails in the distributions, there was a strong influence due to inflow conditions, but overall good agreement between instruments. For the lowest Froude number, excess kurtosis ~ 0, suggesting the data was well represented by a normal distribution with very few extreme data points. In contrast, as Froude numbers increased, excess kurtosis was as high as 7 ($Fr_1$ = 8), suggesting more extreme values. As discussed above, these extreme data points are most likely a result of splashing and ejections by the more violent flow conditions as well as the influence of longitudinal hydraulic jump motions close to the jump toe. Additionally, for the higher Froude numbers, excess kurtosis showed a strong decreasing trend along the length of the hydraulic jump. Overall, there was good agreement between LIDAR and WG (rmsd = 0.8) compared to LIDAR and ADM (rmsd = 1.2) for the measured excess kurtosis.

The characteristic free-surface frequencies were analysed using FFT. The peak of the FFT was selected as the characteristic dominant free surface frequency. In cases with a non-distinct peak, the dominant frequency was determined as the average frequency within a range of 1 Hz before the sharp decay in the frequency spectrum. No secondary frequencies were considered in the present study. Typical FFT distributions are shown for the three instrumentations at an example position along the hydraulic jump in Figure S8. The example FFTs for the LIDAR, ADM and WG show similar distributions with a dominant frequency $F_{fs} \approx 2.0$ Hz. Independent of the measurement position and flow conditions, frequency analysis of the LIDAR and WGs presented more distinct peaks compared to the ADMs (Figure S8). The FFT data for the LIDAR were consistent with previous studies [3,4], while the FFT for the ADMs were comparable to data of Wang and Chanson [5], but had less distinct peaks compared to data presented by Murzyn and Chanson [6]. The differences between Murzyn and Chanson [6] and this study may be due to spot size, but this requires further investigation.

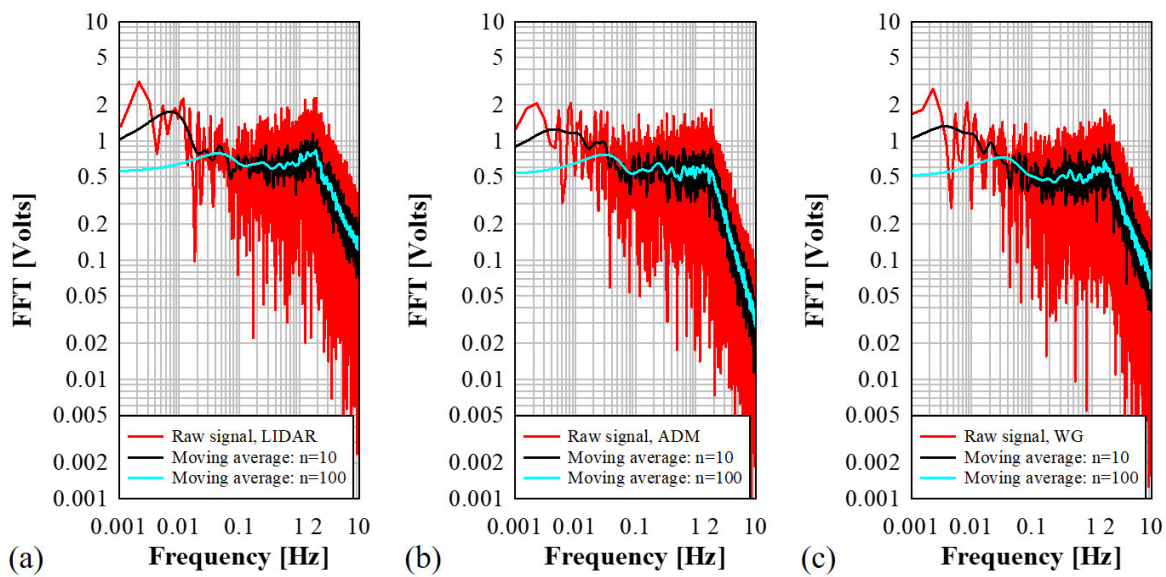

Figure S8. FFT analysis representing typical characteristic free-surface frequencies in aerated hydraulic jumps for $(x-X_{toe,LIDAR})/L_r = 0.39$, $d_1 = 0.028$ m, $Fr_1 = 8$, $Re = 1.2 \times 10^5$: (a) LIDAR; (b) ADM; (c) WG.

Figure S9 shows all distinct and indistinct frequency peaks for LIDAR, ADM and WG data along the jump roller. The frequencies are shown as Strouhal number $F_{fs} \times d_1/v_1$. The magnitude and distributions of characteristic frequencies were similar irrespective of the instrumentation, with $0.2 < F_{fs} < 3.7$ Hz. While characteristic frequencies for all instruments showed some data scatter, the characteristic frequencies close to the jump toe $((x-X_{toe,\,LIDAR})/L_r < 0.2)$ resembled frequencies of the jump toe movement (0.8 – 1 Hz). These findings, as well as a decrease in Strouhal numbers with increasing $Fr_1$ were consistent with previous studies [4–8].

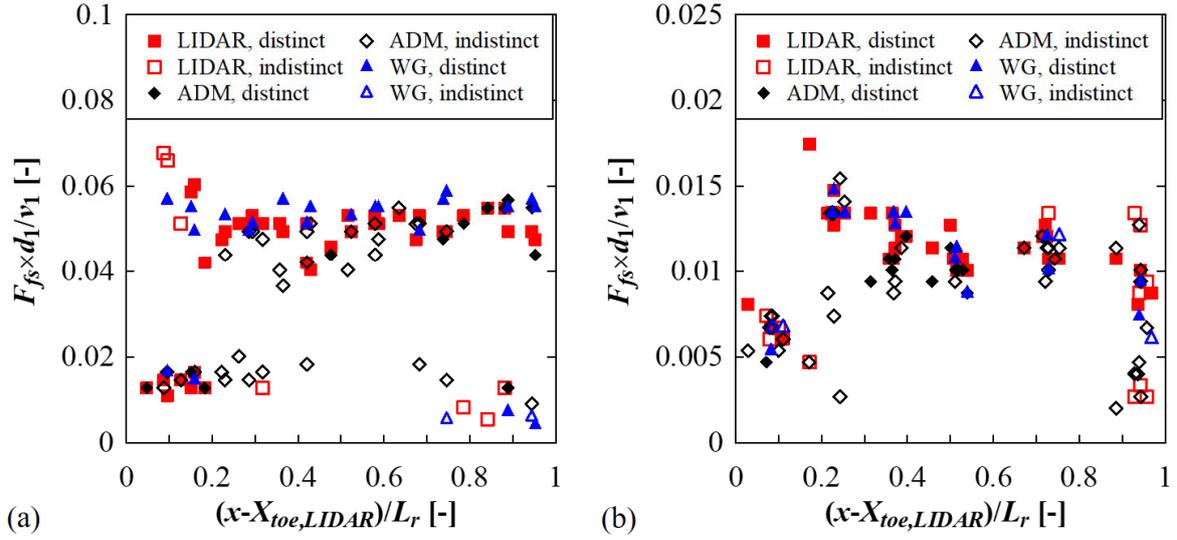

Figure S9. Characteristic dimensionless free-surface frequencies along the jump roller for LIDAR, ADMs and WGs: (a) $d_1 = 0.041$ m, $Fr_1 = 3.5$, $Re = 9.2 \times 10^4$; (b) $d_1 = 0.028$ m, $Fr_1 = 8$, $Re = 1.2 \times 10^5$.

## S4. Comparative analysis of free-surface time scales

This section complements the presentation and discussion of the cross-correlation analysis and free-surface length scales in Section 4 of the main manuscript. To estimate the advective time scales of the free-surface structures, the auto-correlation functions were integrated with the trapezoidal rule until the first crossing of the x-axis or the minimum auto-correlation coefficient if non-zero auto-correlation existed providing the integral auto-correlation time scales $T_{xx}$ [7,8]:

$$T_{xx} = \int_{\tau=0}^{\tau=\tau(R_{xx}=R_{xx,min}\;//\;R_{xx}=0)} R_{xx}(\tau)\,d\tau \qquad (1)$$

where $\tau$ is the time lag, $R_{xx}$ the auto-correlation function and $R_{xx,min}$ the minimum auto-correlation coefficient.

Typical auto-correlation functions for the three instruments are shown in Figure S10 for simultaneously sampled signals comprising $R_{xx}$ for the raw signals (upper row) and the corresponding high-pass filtered (0.1 Hz) signals (bottom row). For the raw signals, the auto-correlation function patterns for $Fr_1$ = 3.5 and 5 were similar and no crossing of the x-axis was observed for any of the instruments for $(x-X_{toe,LIDAR})/L_r < 0.8$ (Figure S10a), while further downstream a crossing of the x-axis occurred (Figure S10b). In the most strongly aerated hydraulic jump with $Fr_1$ = 8 the crossing of the x-axis occurred earlier, i.e. for $(x-X_{toe,LIDAR})/L_r > 0.22$ for LIDAR and WGs (Figure S10b) and for $(x-X_{toe,LIDAR})/L_r > 0.35$ for ADMs (Figure S10c). The high-pass filtering of the signals removed the slow fluctuating component of the data resulting in a consistent downwards shift of the cross-correlation functions and a consistent crossing of the x-axis irrespective of location and instrument (Figure S10).

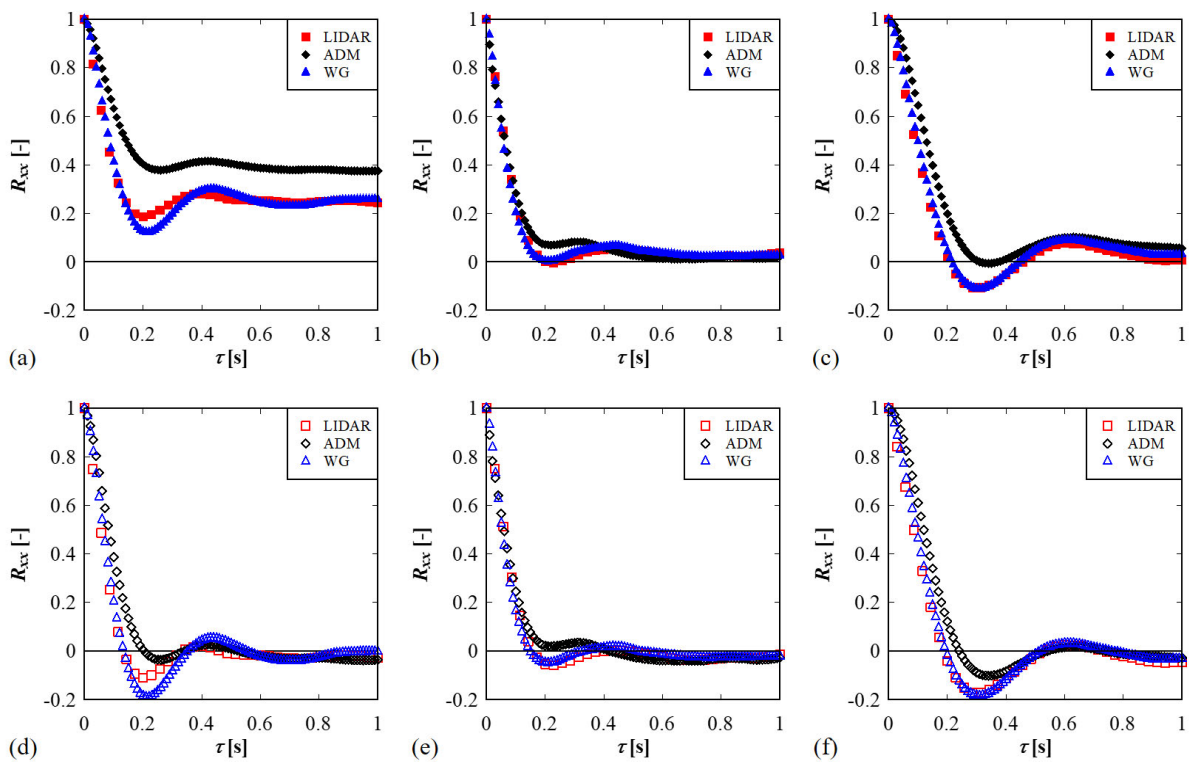

Figure S10. Auto-correlation functions of simultaneously sampled raw free-surface data (top row) and high pass (0.1 Hz) filtered free-surface data (bottom row) in aerated hydraulic jumps with LIDAR, ADM and WG: (a & d) $Fr_1$ = 5, $(x-X_{toe,LIDAR})/L_r$ = 0.32; (b & e) $Fr_1$ = 5, $(x-X_{toe,LIDAR})/L_r$ = 0.92; (c & f) $Fr_1$ = 8, $(x-X_{toe,LIDAR})/L_r$ = 0.51.

The comparison of the auto-correlation functions for the three instruments revealed strong similarity in $R_{xx}$ for the LIDAR and WGs independent of the flow condition and the location along the jump roller. While the overall patterns of the auto-correlation functions were similar for the ADMs, the values of $R_{xx}$ were consistently above the values of the LIDAR and WGs

(Figure S10). This was most pronounced in the first half of the jump roller for the less violent hydraulic jumps with $Fr_1$ = 3.5 and 5 (Figure S10).

It appears that higher auto-correlation determined from the ADM data is based upon two factors: (a) less intensity of the free surface motions including fewer droplet ejections and splashes; and (b) the spot size of the ADM. The hydraulic jumps with the lower Froude numbers were characterised by overall less intense free surface motions in the first part of the jump roller with less intense free surface fluctuations $y'$ and less ejected droplets and spray compared to the hydraulic jump with $Fr_1$ = 8. It appears that a less fragmented free surface provided stronger correlation between the free surface data at a given location resulting in larger auto-correlation functions for $Fr_1$ = 3.5 and 5. The much larger spot size for the ADMs compared to LIDAR and WGs allowed for repeat capture of the free surface motions, possibly including distinct free surface patterns several times leading to higher auto-correlation functions. Additionally, as shown in Figure S8, the ADM did not produce distinct peaks in the characteristic frequency also suggesting some form of smoothing may have occurred due to the larger spot size. This was most pronounced for the hydraulic jumps with lower Froude numbers since the free surface motions were less fragmented compared to the hydraulic jump with $Fr_1$ = 8 and any distinct free surface patterns may be more recognisable in the free surface time series of any instrument.

Figure S11 shows dimensionless integral time scales $T_{xx} \times (g/d_1)^{0.5}$ calculated using the high-pass filtered signal (0.1 Hz) for all instruments. For all data, $T_{xx}$ was analysed starting from $(x-X_{toe,LIDAR})/L_r$, with less than 5% of data being NaNs to eliminate the effect of jump toe motions on the free surface integral scales [8]. Overall, the patterns in $T_{xx}$ were similar for all instruments and flow conditions with slightly lower dimensionless auto-correlation time scales for $Fr_1$ =3.5. The shapes and magnitudes of $T_{xx} \times (g/d_1)^{0.5}$ for the LIDAR data were consistent with the observations of Montano and Felder (2020) with a small peak in $T_{xx} \times (g/d_1)^{0.5}$ at $(x-X_{toe,LIDAR})/L_r \approx 0.7$. While the free surface auto-correlation times scales for LIDAR and WGs were in close agreement with relative differences of less than 18%, $T_{xx} \times (g/d_1)^{0.5}$ measured with the ADMs were consistently larger for $(x-X_{toe,LIDAR})/L_r < 0.8$ (Figure S11). These observations were consistent with observations of the auto-correlation functions (Figure S10) and the lack of distinct characteristic frequencies found by the ADM (Figure S8). As discussed above, increased auto-correlation time scales measured with the ADMs may be a function of spot size, whereby an increased spot size records a stronger connection between the free surface motions and smoothing of the frequencies (Figure S8) leading to a potential overestimation of the true characteristic time scales of the free surface.

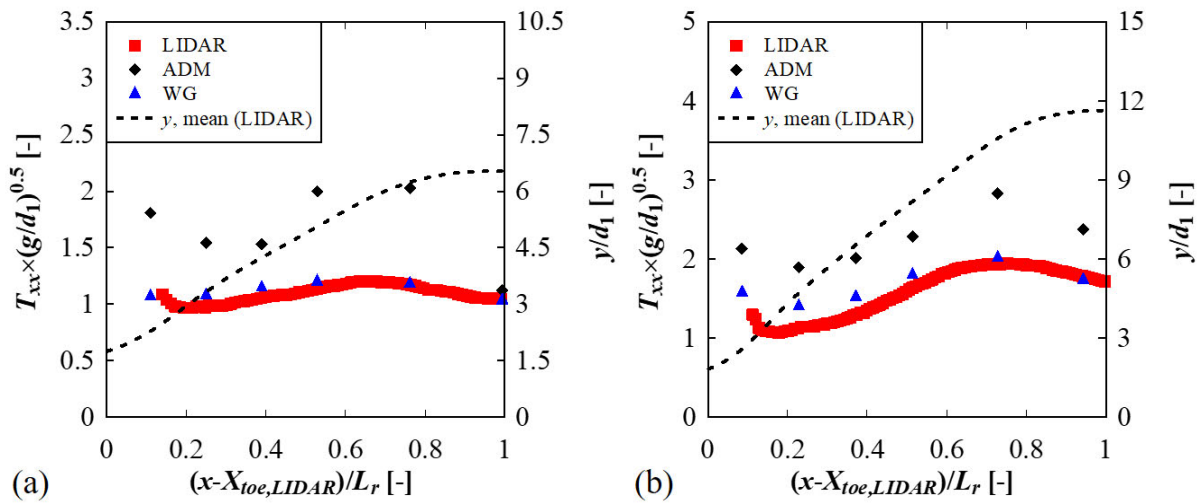

Figure S11. Free-surface integral time scales of high-pass filtered (0.1 Hz) signals in aerated hydraulic jumps measured with LIDAR, ADMs and WGs: (a) $d_1 = 0.034$ m, $Fr_1 = 5$, $Re = 10^4$; (b) $d_1 = 0.028$ m, $Fr_1 = 8$, $Re = 1.2 \times 10^5$.